\documentclass[epj,nopacs]{svjour}
\smartqed  % flush right qed marks, e.g. at end of proof
\usepackage{graphics}
\usepackage{amsmath}
\usepackage{amsfonts}
\usepackage{amssymb}
\usepackage{hyperref}
\usepackage{pdfpages}
\usepackage{cite}
\DeclareGraphicsExtensions{.pdf}
\usepackage{color}

\usepackage{mathtools}
\usepackage{bm} %bold math
\usepackage{ulem}
\usepackage{MnSymbol}

\definecolor{cref2}{rgb}{0.0,0.0,0.0}
\definecolor{cref3}{rgb}{0.0,0.0,0.0}
\definecolor{ccomment}{rgb}{0.0,0.0,0.0}

\newcommand{\Rtwo}[1]{\textcolor{cref2}{#1}}
\newcommand{\Rthree}[1]{\textcolor{cref3}{#1}}

\newcommand{\nullv}{\bm{0}}
\newcommand{\onev}{\bm{1}}

\newcommand{\cv}{\bm{c}} % conformation tensor
\newcommand{\cvh}{\hat{\bm{c}}} % conformation tensor
\newcommand{\ch}{\hat{c}} % conformation tensor
\newcommand{\bv}{\bm{b}} % decomposition of c
\newcommand{\bvh}{\hat{\bm{b}}} % decomposition of c

\newcommand{\rv}{\bm{r}} % vector r
\newcommand{\vv}{\bm{v}} % vector v
\newcommand{\Wv}{\bm{W}} % vector W
\newcommand{\Mv}{\bm{M}} % mobility matrix
\newcommand{\Bv}{\bm{B}} % mobility matrix
\newcommand{\Qv}{\bm{Q}} % dumbbell vector
\newcommand{\Rv}{\bm{R}} % average conformation vector
\newcommand{\nv}{\bm{n}} % orientation vector
\newcommand{\XX}{\mathbf{X}}
\newcommand{\YY}{\mathbf{Y}}
\newcommand{\ZZ}{\mathbf{Z}}
\newcommand{\Vconst}{V_{\mathrm{const}}}
\newcommand{\uvec}[1]{\boldsymbol{\widehat{#1}}}

\newcommand{\kv}{\bm{\kappa}} % (transpose of) velocity gradient

\newcommand{\fr}{{\rm{f}}}

\newcommand{\Tr}{{\rm{T}}}
\newcommand{\dr}{{\rm{d}}}
\newcommand{\dt}{\dr t}

\newcommand{\kBT}{k_{\rm{B}} T}

\newcommand{\bnum}{P}

\newcommand{\bea}{\begin{eqnarray}}
\newcommand{\eea}{\end{eqnarray}}

\newcommand{\sumN}{\sum^N_{\mu=1}}

\newcommand{\mytrace}{{\rm{tr}\,}}

\newcommand{\tr}[1]{{\rm tr}\left(#1\right)}

\newcommand{\mhbracket}[1]{\left\langle #1 \right\rangle^{\textrm{It{\^o}}}_{\dt}}

\newcommand{\Lambdatt}{\bm{\Lambda}^{(4)}}
\newcommand{\Lambdasc}{\Lambda^{(4)}}

\begin{document}

\title{Fluctuating viscoelasticity based on a finite number of dumbbells}
\author{
Markus H{\"u}tter \and 
Peter D. Olmsted \and
Daniel J. Read
}


\institute{
Eindhoven University of Technology, Department of Mechanical Engineering, Polymer Technology, PO Box 513, NL--5600 MB Eindhoven, The Netherlands. \email{m.huetter@tue.nl}
\and
Georgetown University,  Department of Physics, Institute for Soft Matter Synthesis and Metrology, 37th and O Streets, N.W., Washington, D.C. 20057, USA. \email{peter.olmsted@georgetown.edu}
\and
School of Mathematics, University of Leeds, Leeds LS2 9JT, U.K. \email{d.j.read@leeds.ac.uk}
}

\date{To be published in European Physical Journal E, Accepted 2 November 2020}

\abstract{
Two alternative routes are taken to derive, on the basis of the dynamics of a finite number of dumbbells, viscoelasticity in terms of a conformation tensor with fluctuations. The first route is a direct approach using stochastic calculus  only, and it serves as a benchmark for the second route, which is guided by thermodynamic principles. In the latter, the Helmholtz free energy and a generalized relaxation tensor play a key role. It is shown that the results of the two routes agree only if a finite-size contribution to the Helmholtz free energy of the conformation tensor is taken into account. Using statistical mechanics, this finite-size contribution is derived explicitly in this paper for a large class of models; this contribution is non-zero whenever the number of dumbbells in the volume of observation is finite. It is noted that the generalized relaxation tensor for the conformation tensor does not need any finite-size correction.
}

\maketitle

\section{Introduction}

\hrule
\includegraphics[width=0.4\textwidth]{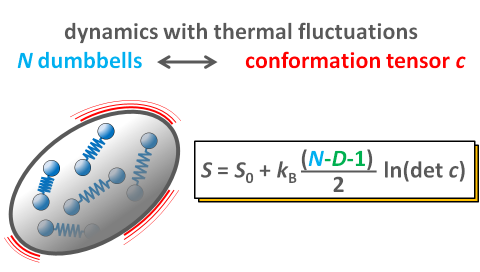}
\hrule
\vskip0.5truecm

Fluctuations are particularly important when studying small systems. This also holds for fluids, including complex fluids, e.g., macromolecular and polymeric liquids. Small scales are involved, e.g., in microrheology \cite{squiresmason2010} and micro- and nanofluidic devices \cite{micro1,micro2}. For Newtonian fluids, i.e., fluids with a deformation-independent viscosity and a lack of memory, the dynamics on small scales could be described in terms of the fluctuating Newtonian fluid dynamics developed by Landau and Lifshitz \cite{LL1959}. However, this is not sufficient for complex fluids, and thus extensions are needed. For example, the stress tensor has been related to the rate-of-strain tensor by a memory kernel, and correspondingly colored noise has been introduced on the stress tensor \cite{hohenegger2017,hohenegger2017addtl}. Another approach towards modeling fluctuating effects in complex fluids has been taken by 
V{\'a}zquez-Quesada, Ellero, and Espa{\~n}ol \cite{vazquezquesadaetal2009} and applied to microrheology \cite{vazquezquesadaetal2012}, in which smoothed-particle hydrodynamics is extended by a conformation tensor that describes the conformation of the small number of polymer chains per volume element. The concept of fluctuating dynamics for the conformation tensor has been extended recently \cite{HHA2018}, to make it applicable not only to the Maxwell model \cite{berisedwards1990b,wapperomhulsen1998}, as in \cite{vazquezquesadaetal2009,vazquezquesadaetal2012}, but to a wider class of models, e.g. the FENE-P model \cite{wapperomhulsen1998,wedgewoodbird1988} and the Giesekus model \cite{wapperomhulsen1998,Giesekus1982RA,Giesekus1982JNNFM,birdwiest1985}. In the approach taken in \cite{HHA2018}, the Helmholtz free energy in terms of the conformation tensor plays an essential role.

The dynamics of the conformation tensor roots in a finer description, in particular, it can be related to the kinetic theory of dumbbells (e.g., see chapter~13 in \cite{birdbook}). The question addressed in this paper is what lessons can be learned from deriving the dynamics for the conformation tensor with fluctuations from an underlying kinetic description for a finite number of dumbbells. \Rtwo{It is pointed out that the dumbbell description already contains the relaxation and fluctuation effects that are relevant also on the conformation-tensor level. This is in contrast to coarse graining from an atomistic description to bead-spring chains or directly to the conformation tensor, e.g., see the work of Underhill and Doyle \cite{underhilldoyle2004} and of Ilg {\it et al.} \cite{ilg2008,ilg2009}, respectively, without fluctuations on the conformation-tensor level.}

The paper is organized as follows. In sect.~\ref{sec:kineticmodels}, a certain class of kinetic dumbbell models is introduced, based on which a description of fluctuating viscoelasticity in terms of the conformation tensor is derived via a direct route, for a finite number of dumbbells. This route is paralleled in sect.~\ref{sec:TDapproach}, where a thermodynamic approach is taken to arrive at fluctuating viscoelasticity. As part of that, the finite-size correction to the Helmholtz free energy is calculated, and this is found to be essential for finding agreement between the two approaches. In the appendix, three different calculation methods for this free energy are detailed, each of which arrives at the same result. In sect.~\ref{sec:multidecomp}, the relation between the dumbbell models and the multiplicative decomposition of the conformation tensor, which has been discussed recently in the literature \cite{HHA2018,publ72,HCHA2020}, is examined. The paper ends with conclusions and a discussion in sect.~\ref{sec:discussion}.

Throughout this paper, the following notation will be used. All summations are spelled out, i.e., no Einstein summation-convention is used for repeated indices. While the symbol $\cdot$ denotes a contraction of one pair of indices, we use $\odot$ for a double contraction: For an order-four tensor $\bm{A}^{(4)}$ and an order-two tensor $\bm{B}$, 
$\left[\bm{A}^{(4)} \odot \bm{B} \right]_{ij} = \sum_{k,l} A^{(4)}_{ijkl}  {B}_{kl}$. The Kronecker delta is given as $\delta_{ij}$, and the Dirac delta-function as $\delta(x-y)$.  The dyadic product of two vectors $\vv_1$ and $\vv_2$ is written as $\vv_1 \vv_2$.

\color{black}

\section{Fluctuating viscoelasticity derived from dumbbell models}  \label{sec:kineticmodels}

\subsection{Dumbbell models} \label{sec:dumbbellmodels}

Let us consider $N$ dumbbells described by connector vectors $\Qv_\mu$, $\mu = 1, \ldots, N$, in $D$-dimensional space. For the purpose of this study, the following class of models is examined (see also \cite{birdbook,hcostochbook,generic2,hcobook}),
\bea
\dr\Qv_\mu &=& \left[\kv\cdot\Qv_\mu - \Mv \cdot \frac{\partial \Phi}{\partial \Qv_\mu} + \kBT \left(\frac{\partial}{\partial \Qv_\mu}\cdot\Mv \right) \right] \dt \nonumber \\
{}&& + \sqrt{2 \kBT} \Bv \cdot \dr \Wv_\mu \, , \label{eq:generalSDE}
\eea
which is a set of coupled stochastic differential equations (SDE) \cite{gardinerbook,hcostochbook}; the It{\^o} interpretation of stochastic calculus will be used throughout this paper \cite{gardinerbook,hcostochbook}. In eq.~\eqref{eq:generalSDE}, 
$\kv = \partial\bm{v}/\partial\rv$ is the gradient of the imposed velocity field, $\Phi$ is the potential energy of all dumbbells, $\Mv$ is the mobility tensor, with $\Mv^\Tr = \Mv$, $T$ stands for the absolute temperature, $\Bv$ must satisfy the \Rtwo{fluctuation-dissipation theorem \cite{Groot,Kubo,Evans,hcobook},} 
\bea
\Bv \cdot \Bv^\Tr = \Mv \, ,  \label{eq:decomposition}
\eea
and the vectors $\dr \Wv_\mu$ are increments of independent Wiener processes \cite{gardinerbook,hcostochbook}, satisfying
\bea
\langle \dr \Wv_\mu (t) \rangle &=& \nullv \, , \quad \forall \mu \, , \\
\langle \dr \Wv_\mu (t) \dr \Wv_\nu (t^\prime) \rangle &=& \delta_{\mu\nu} \, \delta(t - t^\prime) \, dt \, dt^\prime \onev \, , \quad \forall \mu,\nu \, . 
\eea
It is noted that, in general, $\Bv$ has dimensions $D \times P$ with $P \ge D$, and therefore $\dr \Wv_\mu$, for any dumbbell $\mu$, has dimensions $P \times 1$, i.e., is a $P$-dimensional vector for every particle $\mu$. For practical purposes, one may choose $P=D$, although other choices may also be convenient \cite{HHA2018}.

It can be shown that the probability distribution for the system \eqref{eq:generalSDE} at equilibrium is given by the Boltzmann distribution with energy $\Phi$; to ensure that, the third term in the square brackets in eq.~\eqref{eq:generalSDE} is essential  --  if this term was absent, the equilibrium distribution would depend on the mobility tensor, which is unphysical. Note that the mobility $\Mv$ may depend on $\{\Qv_\mu\}_{\mu=1,\ldots,N}$, however, it is assumed in this study that it is the same mobility tensor for all dumbbells $\mu$, and thus it has no subscript $\mu$.

For completeness, it is mentioned that all position vectors are dimensionless, i.e., they are scaled with respect to (w.r.t.) a characteristic constant length-scale, that is omitted throughout this paper for convenience. 

In the following, expressions for $\Phi$ and $\Mv$ are considered that will allow us to eventually derive a closed evolution equation for the instantaneous conformation tensor
\bea
\cvh = \frac{1}{N} \sumN \Qv_\mu \Qv_\mu \equiv \overline{\Qv \Qv} \, . \label{eq:ctensor}
\eea
Symmetry requires that the potential energy $\Phi_{\rm d}$ of a single dumbbell depends only on the (squared) length of the dumbbell vector. If the dumbbells are not interacting, the total potential energy of all dumbbells is given by the sum of the individual dumbbell contributions, i.e.,  
$\Phi_{\rm exact} = N \overline{ \Phi_{\rm d}(\tr{\Qv \Qv}) }$. However, in this paper, we consider models for which the total potential energy is obtained by interchanging the operations $\overline{ (\ldots) }$ and $\Phi_{\rm d}(\tr{\ldots})$ in $\Phi_{\rm exact}$, i.e., we use the relation
\bea
\Phi = N \Phi_{\rm d}(\mytrace{\cvh}) \, , \label{eq:approxphi}
\eea
where eq.~\eqref{eq:ctensor} was employed. This mean-field ansatz will be beneficial for deriving closed dynamics for $\cvh$. 
\Rtwo{Obviously, eq.~\eqref{eq:approxphi} is --- for all but one case (see below) --- only an approximation to the exact potential for all dumbbells. However, for the purpose of this paper (which is to examine the effect of finite $N$ on the counting of states, and ramifications thereof for the dynamics of $\cvh$), we employ eq.~\eqref{eq:approxphi} for defining the class of models we examine. Therefore, we will not go into details about how accurately eq.~\eqref{eq:approxphi} approximates the exact potential. }

The energy $\Phi$ given by eq.~\eqref{eq:approxphi} equals the exact energy $\Phi_{\rm exact}$ only if the function $\Phi_{\rm d}(\tr{\ldots})$ is linear, in particular for linear-elastic Hookean springs, $\Phi_{\rm d}(\tr{\Qv \Qv}) = (H/2) \tr{\Qv \Qv}$, with $H$ the spring constant of the dumbbell. However, other cases are of interest as well.
The dumbbell force generally has the form $ - H f \Qv_\mu$. For example, for a linear-elastic Hookean spring, $f=1$, while for a finitely extensible nonlinear elastic spring one can use \footnote{Sometimes a different parametrization is used, e.g. $\beta_1 = (b+3)/b$ and $\beta_2 = b H / \kBT$, with dimensionless parameter $b$ \cite{HHA2018}.} (see also  \cite{wedgewoodbird1988}),
\bea
f = \left( \beta_1 - \frac{H}{\beta_2 \kBT}\mytrace{\cvh} \right)^{-1} \, , \label{eq:FENEP:f}
\eea
with two constants $\beta_1$ and $\beta_2$, and 
which accounts for the finite extensibility in a mean-field sense. Typically, in the FENE-P approximation \cite{wedgewoodbird1988} (see also \cite{peterlin1966}), $\cvh$ in eq.~\eqref{eq:FENEP:f} is used for infinitely many dumbbells. However, since in this paper the focus is on studying systems with a finite number $N$ of dumbbells, we generalize this by using eq.~\eqref{eq:FENEP:f} for finite $N$. The potential that corresponds to the desired expression for the force is given by 
\bea
\Phi = \frac{N \beta_2 \kBT}{2} \ln f 
%\quad \xrightarrow[\beta_1 = 1, \beta_2 \to \infty]{} \quad \frac{N H}{2} \mytrace{\cvh} 
\, ,  \label{eq:FENEP:phi} 
\eea
of which the limit $\beta_1 = 1$ and $\beta_2 \to \infty$ results in the potential for the linear-elastic Hookean spring. 
Therefore, using the approximation $\Phi$, given by eq.~\eqref{eq:FENEP:phi} with eq.~\eqref{eq:FENEP:f}, 
instead of the exact energy $\Phi_{\rm exact}$ is safe for small deformations. 
\Rtwo{For other deformations, it is noted that the potential $\Phi$ does include finite extensibility, albeit in a different way than if applied to each dumbbell individually.}
It is a topic of future research to examine the foundations of the mean-field approximation $\Phi$ for finite $N$ thoroughly. In this paper, this approximate expression for the energy forms part of defining the kinetic models that are examined in the following, and it is suitable for deriving closed dynamics for $\cv$, in the same spirit as the FENE-P approximation has been introduced earlier for infinite $N$ \cite{wedgewoodbird1988}. 

As far as the mobility tensor $\Mv$ is concerned, 
its dependence on the dumbbell vectors $\{\Qv_\mu\}_{\mu = 1, \ldots, N}$ is restrict\-ed to a dependence on the instantaneous conformation tensor,
\bea
\Mv = \Mv(\cvh) \, . \label{eq:general:M}
\eea
A particular realization of that is
\bea
\Mv &=& \frac{2}{\zeta} \left((1-\alpha) \onev + \alpha \cvh \right)\, , \label{eq:Giesekus:M}
\eea
with friction coefficient $\zeta$, and the parameter $\alpha$ is used to adjust the amount of $\cvh$-dependence. In case of imposed deformation, a non-zero value for $\alpha$ thus results in anisotropy of the friction tensor, where the 
anisotropy is introduced in a mean-field sense. The form \eqref{eq:Giesekus:M} corresponds to the widely used Giesekus model for anisotropic drag \cite{Giesekus1982RA,Giesekus1982JNNFM,birdwiest1985}. 
For the Giesekus model, one typically uses eq.~\eqref{eq:Giesekus:M} for infinitely many chains, i.e., $N \to \infty$, to render the model solvable. However, the mobility of a dumbbell is affected primarily by the other dumbbells in its vicinity, and therefore the finite-$N$ generalization \eqref{eq:Giesekus:M} is reasonable. 

Allowing for the dumbbell potential energy $\Phi_{\rm d}(\tr{\ldots})$ in eq.~\eqref{eq:approxphi} to be nonlinear and/or the dumbbell mobility tensor $\Mv$ to depend on the conformation tensor effectively introduces mean-field type couplings of the individual dumbbells. In practice this implies that either (a) the $N$ dumbbells must be in the vicinity of each other so they can interact, or (b) they diffuse rapidly enough in space to effect such interactions. In the case of the mobility tensor, the implied physics is reasonably clear: the assumption is that the average orientation of surrounding dumbbells affects the mobility of any given test dumbbell.  Having this rationale in mind suggests that this mean-field mobility makes more sense for a finite number $N$ of dumbbells, than it does for infinitely many. In the case of the potential energy, the microscopic physics of the implied coupling between dumbbells is less clear.  Still, we find it worthy of note that a potential energy function exists from which the FENE-P model can be derived exactly.

In order to highlight the overall structure of the modeling in the remainder of this paper, the general forms eq.~\eqref{eq:approxphi} for the potential $\Phi$ and eq.~\eqref{eq:general:M} for the mobility tensor $\Mv$ will be used. These general results can then be reduced to the Hookean dumbbell model, the FENE-P model, and the Giesekus model, respectively, by appropriate choices for the forms and parameters for the potential $\Phi$ and the mobility tensor $\Mv$, see table~\ref{tab:models}.

\begin{table}[t]
\centering
\caption{Overview of the parameters in the potential $\Phi$ given by eq.~\eqref{eq:FENEP:f} and eq.~\eqref{eq:FENEP:phi}, and the mobility tensor $\Mv$, eq.~\eqref{eq:Giesekus:M}, for the three models}
\label{tab:models}
\vspace{0.5cm}
\begin{tabular}{l|cc}
\hline\noalign{\smallskip}
\multicolumn{1}{l|}{Dumbbell} & \multicolumn{2}{l}{Parameters:} \\
 model & $\Phi$ & $\Mv$ \\
\noalign{\smallskip}\hline\noalign{\smallskip}
Hookean & $\beta_1 = 1$, $\beta_2 \to \infty$  ($f = 1$) & $\alpha = 0$ \\
FENE-P & $\beta_1 = 1$, $\beta_2$ finite ($f \neq 1$) & $\alpha = 0$ \\
Giesekus & $\beta_1 = 1$, $\beta_2 \to \infty$ ($f = 1$) & $\alpha > 0$ \\
\noalign{\smallskip}\hline 
\end{tabular}
\vspace{1.0cm}
\end{table}

\subsection{Transition from dumbbells to the conformation tensor} \label{sec:rtoc}

Given the definition of the instantaneous conformation tensor $\cvh$, eq.~\eqref{eq:ctensor}, and using the It{\^o} interpretation of stoch\-astic calculus  \cite{McKean,hcostochbook}, one has in general
\bea
\dr \cvh = \frac{1}{N} \sumN \left(
(\dr \Qv_\mu) \Qv_\mu + \Qv_\mu (\dr \Qv_\mu) + \mhbracket{(\dr \Qv_\mu) (\dr \Qv_\mu)}
\right) \, , \nonumber \\\label{eq:chainrule}
\eea
where $\mhbracket{\ldots}$ implies that in $\dr \Qv_\mu$ only terms involving the Wiener increments are kept and subsequently reduced according to the rule (see Table~3.1 in \cite{hcostochbook})
\bea
\dr \Wv_\mu \dr \Wv_\mu  &\rightarrow& \dt \onev \, . \label{eq:McKean2} 
\eea
Applied to the above class of models, the SDEs \eqref{eq:generalSDE} for the dumbbell vectors $\{\Qv_\mu\}_{\mu=1,\ldots,N}$ can be transformed into an SDE for the conformation tensor,
%\color{cdelete}
%\bea
%\dr \cvh &=& \left[ \kv\cdot\cvh + \cvh\cdot\kv^\Tr 
%- \frac{4 H}{\zeta} \left((1-\alpha) \onev + \alpha \cvh \right) \cdot \left(f \cvh - \frac{\kBT}{H} \onev \right) \right] \dt \nonumber \\
%{}&&
%+\frac{4 \alpha (D+1) \kBT}{N \zeta} \cvh \dt + \dr \cvh^\fr \, , \label{OLDeq:SDE:c:general}
%\eea
%\color{black}
\bea
\dr \cvh &=& \left[ \kv\cdot\cvh + \cvh\cdot\kv^\Tr 
- 2 \Mv \cdot \left( 2 \frac{\partial \Phi_{\rm d}}{\partial (\mytrace{\cvh})} \cvh 
- \kBT \onev \right) \right] \dt \nonumber \\
{}&&
+ \frac{2 \kBT}{N} \left[ 
\cvh \cdot \frac{\partial}{\partial \cvh} \cdot \Mv 
+
\left( \cvh \cdot \frac{\partial}{\partial \cvh} \cdot \Mv \right)^\Tr
\right] \dt + \dr \cvh^\fr \, , \label{eq:SDE:c:general}
\eea
in terms of the dumbbell potential energy $\Phi_{\rm d}$ and the dumbbell mobility tensor $\Mv$,
where it has been assumed that $\Mv\cdot\cvh = \cvh\cdot\Mv$. It is pointed out that the symmetry of $\cvh$ must be taken into account explicitly when calculating the partial derivatives of $\Mv$ (see \cite{HHA2018} for details).
The quantity $\dr \cvh^\fr$ denotes the thermal fluctuations,
\bea
\dr \cvh^\fr = 
\frac{\sqrt{2 \kBT}}{N} \sumN \left( \left(\Bv \cdot \dr \Wv_\mu\right)  \Qv_\mu 
+ \Qv_\mu \left(\Bv \cdot \dr \Wv_\mu\right) \right) \, , \label{eq:noise:c:general}
\eea
with $\Bv$ given by eq.~\eqref{eq:decomposition} for a general mobility tensor $\Mv$, eq.~\eqref{eq:general:M}. It can be shown that the fluctuations have the properties
\bea
\langle \dr \ch^\fr_{ij} \rangle &=& 0 \, , \label{eq:Giesekus:noise:avg} \\
\langle \dr \ch^\fr_{ij} \dr \ch^\fr_{kl} \rangle &=& 
\frac{2 \kBT}{N} \left(
\ch_{ik} M_{jl} + \ch_{il} M_{jk} \right. \nonumber \\
{}&& \qquad \quad \left. + \ch_{jk} M_{il} + \ch_{jl} M_{ik} 
\right) \dt \, . \label{eq:Giesekus:noise:cov} 
\eea

The SDE \eqref{eq:SDE:c:general} for the conformation tensor with fluctuations obeying the statistical properties given by eq.~\eqref{eq:Giesekus:noise:avg} and eq.~\eqref{eq:Giesekus:noise:cov} is the benchmark to which the thermodynamic treatment further below will be compared, for the general class of models described by energy $\Phi$, eq.~\eqref{eq:approxphi}, and mobility tensor $\Mv$, eq.~\eqref{eq:general:M}. \Rtwo{If the potential $\Phi$ and the mobility tensor $\Mv$ are of forms more general than eq.~\eqref{eq:approxphi} and eq.~\eqref{eq:general:M}, respectively, the dynamics for the conformation tensor would not close automatically, in which case one would have to employ procedures of coarse graining \cite{hcoCG,hcobook,espanol2002,espanol2009}.} The procedure presented in this sec.~\ref{sec:rtoc} has also been followed in \cite{vazquezquesadaetal2009} for deriving the $\cvh$-dynamics for the Hookean and FENE-P models.

It is noted that, in the limit of many dumbbells, i.e., $N \to \infty$, not only do the fluctuations $\dr \cvh^\fr$
become insignificant. In addition, also the ``thermal drift'', i.e., the second-last  contribution on the right-hand side (r.h.s.) of eq.~\eqref{eq:SDE:c:general} vanishes as well, and thus the conventional deterministic dynamics for the conformation tensor is recovered. 

The Hookean dumbbell model  --  which results in the Maxwell model for the conformation tensor  --  , the FENE-P model, and the Giesekus model are sub-cases of the SDE \eqref{eq:SDE:c:general} with noise \eqref{eq:noise:c:general} when choosing the parameters $\beta_1$, $\beta_2$, and $\alpha$ appropriately, see table~\ref{tab:models}. 

\section{Fluctuating viscoelasticity derived using thermodynamics} \label{sec:TDapproach}

The main idea in taking a thermodynamic approach to modeling dynamical systems is that one can concentrate on key ingredients for the static and dynamic properties, and the thermodynamic approach makes sure that these ingredients are processed towards the final model in a consistent way. For the dumbbell dynamics \eqref{eq:generalSDE}, the key ingredients are the potential $\Phi$ and the mobility tensor $\Mv$. In contrast, for the dynamics of $\cv$, the key ingredients are the Helmholtz free energy density $\psi$ and the generalized relaxation tensor $\Lambdatt$ (see \cite{HHA2018} for details). In particular, the dynamics for the fluctuating conformation-tensor $\cv$ is given by the SDE  
\bea
\dr\cv &=& \left(\kv\cdot\cv + \cv\cdot\kv^{\Tr} - \Lambdatt \odot \frac{\partial \psi}{\partial \cv} \right) \dr t \nonumber \\
{}&&+ \frac{\kBT}{V} {\rm div}_{\cv} \Lambdatt \dr t
+ \sqrt{\frac{2 \kBT}{V}} \bm{B}^{(4)} \odot \dr \tilde{\Wv} \, , \label{eq:cflu}
\end{eqnarray}
where ${\rm div}_{\cv} \Lambdatt$ denotes the divergence of $\Lambdatt$ in $\cv$-space, the order-four tensor $\bm{B}^{(4)}$ satisfies 
\bea
\bm{B}^{(4)} \odot \bm{B}^{(4), \Tr} = \Lambdatt \, , \label{eq:p65:FDT}
\eea
and $\dr \tilde{\Wv}$ is a tensor with increments of independent Wiener processes (see \cite{HHA2018} for further details). The structure of the SDE \eqref{eq:cflu} with eq.~\eqref{eq:p65:FDT} is completely analogous to the one for the dumbbell models, eq.~\eqref{eq:generalSDE} with eq.~\eqref{eq:decomposition}.

In the following, we derive expressions for $\psi$ and $\Lambdatt$, based on those for $\Phi$ and $\Mv$, respectively. 

\subsection{Free energy density $\psi(\cv)$ for finite $N$} \label{sec:freeenergy}

The Helmholtz free energy $\Psi = \Psi(\cv)$ for the symmetric conformation-tensor $\cv$ for a finite number $N$ of dumbbells is given by $\Psi = -\kBT \ln Z$, with the canonical partition-function
\bea
Z(\cv) = \int e^{-\Phi(\cvh)/(\kBT)} \, \delta^{(K)}\left(\cvh-\cv\right) \, d^{D N}Q \, , \label{eq:Z:full}
\eea
where $D$ is the number of spatial dimensions. The $K$-dimensional Dirac $\delta$-function makes sure that only those states in $\{\Qv_\mu\}$-space are accounted for that are compatible with the conformation tensor $\cv$. Since $\cvh$ is symmetric by definition, see eq.~\eqref{eq:ctensor}, only $K = D(D + 1)/2$ independent conditions are needed (instead of $D^2$); no more conditions
are required for properly restricting the integration in $\{\Qv_\mu\}$-space.

It is pointed out that $\delta^{(K)}$ is actually a $\delta$-function in $\cv$-space in the sense that $\int \delta^{(K)}(\cvh-\cv) d^{K}c = 1$. Using this latter relation, the integral of $Z(\cv)$ over all (symmetric) conformation tensors $\cv$ reduces to 
\bea
\Xi &=& \int Z(\cv) d^Kc = \int e^{-\Phi(\cvh)/(\kBT)} \, d^{D N}Q \, , \label{eq:Xi:full}
\eea
which is the conventional canonical partition function in the absence of the constraint $\cvh=\cv$.

As we restrict our attention to energy functions $\Phi$ which depend on $\{\Qv_\mu\}_{\mu=1,\ldots,N}$ only by way of $\cvh$,
see eq.~\eqref{eq:approxphi}, the canonical partition-function $Z$ is related to the microcanonical partition-funct\-ion $\Gamma$ by way of
\bea
Z(\cv) = e^{-\Phi(\cv)/(\kBT)} \, \Gamma(\cv) \, , \label{eq:Z:factors}
\eea
with
\bea
\Gamma(\cv) = \int \, \delta^{(K)}\left(\cvh-\cv\right) \, d^{DN}Q \, . \label{eq:Gamma:general}
\eea
Different procedures for calculating the dependence of $\Gamma$ on $\cv$ explicitly are discussed in Appendix~\ref{sec:appendix1}, one based on deriving a differential equation for $\Gamma$, another one with a more geometrical interpretation, and a third one using a scaling argument. Following any of these procedures, the result for finite $N$ is
\bea
\Gamma = \Gamma_0 \left(\det\cv\right)^{(N-D-1)/2} \, , \label{eq:Gamma:result}
\eea
with a $\cv$-independent prefactor $\Gamma_0$. 

Based on eq.~\eqref{eq:Z:factors} with eq.~\eqref{eq:Gamma:result}, the Helmholtz free energy density $\psi = \Psi/V$ per volume $V$ becomes
\bea
\psi = 
\frac{\Phi}{V} 
-\frac{n \kBT}{2} \ln \left(\det\cv\right) 
+ \Delta\psi \, , \label{eq:psi}
\eea
with
\bea
\Delta\psi = 
\frac{\kBT}{2 V} (D+1) \ln \left(\det\cv\right) \, , \label{eq:deltapsi}
\eea
and number density $n = N/V$, and where we have omitted a $\cv$-independent additive constant, which is irrelevant for the formulation of the dynamics of $\cv$ according to eq.~\eqref{eq:cflu}. Since $\Phi$ is proportional to $N$, it is evident that the first two contributions on the r.h.s. of eq.~\eqref{eq:psi} are independent of the size of the system, for given number density $n$. In contrast, the third contribution, $\Delta\psi$, does depend on the size of the system, in particular it becomes more relevant the smaller the system. To the best of our knowledge, this finite-size correction to the Helmholtz free energy (density) has not been derived earlier.

Using $H = \kBT$ and $G = n \kBT$, the Helmholtz free energy density \eqref{eq:psi} with eq.~\eqref{eq:deltapsi} for the three models discussed in table~\ref{tab:models} agrees with standard literature (e.g., see \cite{birdbook,berisedwards1990b,wapperomhulsen1998}), and with what has been used in the fluctuating-viscoelasticity approach in \cite{HHA2018}, with the important difference of the finite-size correction $\Delta\psi$. Using, in contrast to our procedure, a continuous (representative of $N \to \infty$) distribution for the dumbbell vector $\Qv$ (e.g., see \cite{birdbook}), the thereby-derived Helmholtz free energy density corresponds to eq.~\eqref{eq:psi} where the finite-size correction $\Delta\psi$ is absent.

\subsection{Relaxation tensor $\Lambdatt$}

In the dumbbell dynamics \eqref{eq:generalSDE}, structural relaxation is expressed as $-\Mv\cdot\left(\partial \Phi/\partial \Qv_\mu\right) \dt$. In the $\cv_\mu$-dynamics \eqref{eq:cflu}, structural relaxation is expressed as $-\Lambdatt \odot \left(\partial \psi/\partial \cv\right) \dt$, with an order-four relaxation tensor $\Lambdatt$ \cite{HHA2018}. In the translation from $\Mv$ to $\Lambdatt$, a reduction of variables and the volume of the system $V$ are involved, the latter being necessary since $\Phi$ is an energy while $\psi$ is an energy density. The relation between $\Mv$ and $\Lambdatt$ is given by (see also \cite{hcoCG}, and sect. 6.4 in \cite{hcobook})
\bea
\Lambdatt = V \left\langle \sumN \frac{\partial \cvh}{\partial \Qv_\mu} \cdot \Mv \cdot \frac{\partial \cvh}{\partial \Qv_\mu} \right\rangle \, , 
\eea
where the contractions run over the components of $\Qv_\mu$ and $\Mv$, and $\langle\ldots\rangle$ is the average over $\{\Qv_\mu\}$-space for given $\cv$. Using
$\partial\ch_{ij}/\partial Q_{\mu,m} = (\delta_{im} Q_{\mu,j}+\delta_{jm} Q_{\mu,i})/N$, one finds
\bea
\Lambdasc_{ijkl} = \frac{V}{N} \left\langle  
  \ch _{jl} M_{ik} + \ch _{jk} M_{il} + \ch _{il} M_{jk} + \ch _{ik} M_{jl} 
\right\rangle \, . \label{eq:Lambda:general}
\eea
Since $\Mv$ depends on the positions $\{\Qv_\mu\}_{\mu=1,\ldots,N}$ only by way of $\cvh$, taking the average is thus equivalent to replacing $\cvh$ by $\cv$ everywhere in eq.~\eqref{eq:Lambda:general}. It is to be noted that there is {\it no} finite-size correction in $\Lambdatt$.

Using again $G = n\kBT$ and with $\zeta = 4 \kBT \lambda$, the relaxation tensor $\Lambdatt$ given by eq.~\eqref{eq:Lambda:general} for the three models in table~\ref{tab:models} turns out to agree with the standard expressions in the literature  (e.g., see \cite{birdbook,berisedwards1990b,wapperomhulsen1998}) , and with what has been used in \cite{HHA2018} in the context of fluctuating viscoelasticity. 

\subsection{Application to fluctuating viscoelasticity} \label{subsec:applFVE}

According to the general procedure in \cite{HHA2018}, represented in eq.~\eqref{eq:cflu}, and using the Helmholtz free energy density \eqref{eq:psi} with eq.~\eqref{eq:deltapsi} and the relaxation tensor \eqref{eq:Lambda:general} with $\cvh \to \cv$,  one observes that the results in \cite{HHA2018} need to be amended by including the finite-$N$ contribution \Rtwo{
\begin{subequations}
\bea\left. \dr \cv \right|_{\Delta\psi} &=& -\Lambdatt \odot \frac{\partial(\Delta\psi)}{ \partial\cv} \dt  \\ 
{}&=& -2 \kBT ((D+1)/N) \Mv \dr t \, .
\eea
\end{subequations}}
In particular, one obtains for the complete fluctuating dynamics
%\color{cdelete}
%\bea
%\dr \cv &=& \left[\kv\cdot\cv + \cv\cdot\kv^{\Tr} 
%%\right. \nonumber \\ {}&&\left. 
%- \frac{4 H}{\zeta} ( (1-\alpha)\onev + \alpha\cv ) \cdot \left(f \cv - \frac{\kBT}{H}\onev\right)
% \right] \dr t \nonumber \\
%{}&&+ 
%\frac{4 \alpha (D+1) \kBT}{N \zeta} \cv \dr t  + \dr \cv^\fr \, ,  \label{OLDeq:SDEp65}
%\eea
%\color{black}
\bea
\dr \cv &=& \left[\kv\cdot\cv + \cv\cdot\kv^{\Tr} 
%\right. \nonumber \\ {}&&\left. 
- 2 \Mv \cdot \left(2 \frac{\partial \Phi_{\rm d}}{\partial (\mytrace{\cv})}\cv - \kBT \onev
\right)
 \right] \dr t \nonumber \\
{}&&
+ \frac{2 \kBT}{N} \left[ 
\cvh \cdot \frac{\partial}{\partial \cvh} \cdot \Mv 
+
\left( \cvh \cdot \frac{\partial}{\partial \cvh} \cdot \Mv \right)^\Tr
\right] \dt
 + \dr \cv^\fr \, ,  \label{eq:SDEp65}
\eea
 where the symmetry of $\cv$ has been taken into account when calculating ${\rm div}_{\cv} \Lambdatt$ (see \cite{HHA2018} for details). See Appendix~\ref{sec:appendix0} for explicit exemplary applications of this equation.
The symbol $\dr\cv^\fr$ denotes the fluctuating contribution given by
%\color{cdelete}
%\bea
%\dr\cv^\fr &=& 
%\sqrt{\frac{4 \kBT}{N \zeta}} 
%\left[
%\sqrt{1-\alpha}\left(\bv \cdot \dr\tilde{\Wv}_{(1)} + \dr\tilde{\Wv}_{(1)}^{\Tr} \cdot \bv^{\Tr}\right) \right. \nonumber \\
%{}&&\left.\qquad \ \ + 
%\sqrt{\alpha} \bv \cdot \left(
%\dr\tilde{\Wv}_{(2)}  + \dr\tilde{\Wv}_{(2)}^{\Tr} 
%\right) \cdot \bv^{\Tr}
%\right]
% \, . \label{OLDeq:noisep65}
%\eea
%Here, $\bv$, $\dr\tilde{\Wv}_{(1)}$, and $\dr\tilde{\Wv}_{(2)}$ are all $D \times D$, $\bv$ satisfies the condition $\cv = \bv \cdot \bv^\Tr$, and the order-two tensors $\dr\tilde{\Wv}_{(1)}$ and $\dr\tilde{\Wv}_{(2)}$ are increments of statistically independent Wiener processes, each of them having the following properties, 
%\begin{eqnarray}
%\langle \dr\tilde{\Wv}_{(i)}(t) \rangle &=& \nullv \, , \label{OLDeq:noiseavgp65} \\
%\langle \dr\tilde{\Wv}_{(i)}(t) \, \dr\tilde{\Wv}_{(i)}(t^\prime) \rangle &=& \delta(t - t^\prime) \, dt \, dt^\prime \onev^{(4)} \, . \label{OLDeq:noisecorrp65}
%\end{eqnarray}
%\color{black}
\bea
\dr\cv^\fr &=& 
\sqrt{\frac{2 \kBT}{N}} 
\left(
\bv \cdot \dr\tilde{\Wv} \cdot \Bv^\Tr
+
\Bv \cdot \dr\tilde{\Wv}^\Tr \cdot \bv^\Tr
\right)
 \, , \label{eq:noisep65}
\eea
where $\bv$ satisfies the condition 
\bea
\cv = \bv \cdot \bv^\Tr \, . \label{eq:cdecomp:regular}
\eea
In general, $\Bv$ in eq.~\eqref{eq:noisep65}  has dimensions $D \times P$ with $P \ge D$ (as described in sect.~\ref{sec:dumbbellmodels}), $\bv$ has dimensions $D \times P^\prime$ with $P^\prime \ge D$, and therefore $\dr\tilde{\Wv}$ has dimensions $P^\prime \times P$; for practical purposes, one may choose $P = P^\prime = D$. 
The tensor $\dr\tilde{\Wv}$ consists of increments of statistically independent Wiener processes, with the properties 
\begin{eqnarray}
\langle \dr\tilde{W}_{ij}(t) \rangle &=& 0 \, , \label{eq:noiseavgp65} \\
\langle \dr\tilde{W}_{ij}(t) \, \dr\tilde{W}_{kl}(t^\prime) \rangle &=& \delta_{ik} \, \delta_{jl} \, \delta(t - t^\prime) \, dt \, dt^\prime \, . \label{eq:noisecorrp65}
\end{eqnarray}
When using component notation, eq.~\eqref{eq:noisecorrp65} implies that any two of the components of $\dr\tilde{\Wv}$ are independent from each other.

A direct comparison shows that the SDE \eqref{eq:SDE:c:general} derived directly from the dumbbell model and the SDE \eqref{eq:SDEp65} derived via the thermodynamic route, respectively, agree. It is noted that the expressions for the fluctuations, $\dr\cvh^\fr$ in eq.~\eqref{eq:noise:c:general} and
$\dr\cv^\fr$ in eq.~\eqref{eq:noisep65}, respectively, have a different form. However, it can be shown that they have the same statistical properties. First, both representations are linear superpositions of increments of Wiener processes, and second, for both representations the average is given by eq.~\eqref{eq:Giesekus:noise:avg} and the covariance by \eqref{eq:Giesekus:noise:cov}. The difference in the expressions for the fluctuations is not a short-coming of the approach; it rather reflects the non-uniqueness of the decompositions \eqref{eq:p65:FDT} 
and \eqref{eq:cdecomp:regular}. The non-uniqueness of the decomposition \eqref{eq:cdecomp:regular} can actually be utilized for relating the expressions \eqref{eq:noise:c:general} and \eqref{eq:noisep65} for the noise in even more explicit terms. Specifically, choosing $\bv$ to be $D \times N$ with the column vectors of $\bv$ equal to $\Qv_\mu/\sqrt{N}$ ($\mu = 1, \ldots, N$) (see sect.~\ref{sec:multidecomp} for a further elaboration), and setting the row vectors of $\dr\tilde{\Wv}$ equal to $\dr\Wv_\mu$  ($\mu = 1, \ldots, N$), one finds that the expressions \eqref{eq:noise:c:general} and \eqref{eq:noisep65} are identical.

In deriving the relaxation term in eq.~\eqref{eq:SDEp65}, i.e. the first term on the r.h.s. that is proportional to $\Mv$, from the general form eq.~\eqref{eq:cflu}, one notices the following: The prefactor $1/N$ in $\Lambdatt$ given by  eq.~\eqref{eq:Lambda:general} is cancelled by the prefactor $N$ in the derivative $\partial (\psi - \Delta\psi) / \partial \cv$, for $\psi$ given by eq.~\eqref{eq:psi} with $\Phi$ according to eq.~\eqref{eq:approxphi}. If one chose to not cancel these factors, one would observe that $(1/N) \Mv$ (with factor $1/N$) is the relevant mobility on the conformation-tensor level not only for the thermal drift and the fluctuations (by way of the covariances \eqref{eq:Giesekus:noise:cov}), but also for the relaxation.

The importance of the finite-size correction $\Delta \psi$ in the free energy density, eq.~\eqref{eq:deltapsi}, for the evolution equation \eqref{eq:SDEp65} is pointed out. In particular, $\left. \dr \cv \right|_{\Delta\psi} $ exactly cancels those contributions from ${\rm div}_{\cv} \Lambdatt$ that are related to the derivative of the explicit factors $\cvh$ in $\Lambdatt$, eq.~\eqref{eq:Lambda:general}. If $\Delta \psi$ was neglected, agreement between the SDEs \eqref{eq:SDE:c:general} and \eqref{eq:SDEp65} could not be achieved.

In Appendix~\ref{sec:appendix0}, the dynamics for the conformation tensor with fluctuations, eq.~\eqref{eq:SDEp65} with eq.~\eqref{eq:noisep65}, is presented explicitly for three models, namely for the Hookean dumbbell / Maxwell model, the FENE-P model, and the Giesekus model.

%\color{cdelete}
%It is pointed out that for the agreement between the SDEs \eqref{eq:SDE:c:general} and \eqref{eq:SDEp65}, it is essential that the finite-size correction in the Helmholtz free energy density, i.e., $\Delta \psi$ given by eq.~\eqref{eq:deltapsi}, is taken into account. If this contribution to $\psi$ were omitted, the thermal drift (the contribution just before $\dr\cv^\fr$ in eq.~\eqref{eq:SDEp65}) would have a different prefactor, which would result in erroneous behavior of the model, e.g., in an increased magnitude of $\cv$ in equilibrium for the Maxwell model.
%\color{black}

\section{Comments on the multiplicative decomposition of c} \label{sec:multidecomp}

\subsection{Eliminating degrees of freedom}
Above, the relation has been established between the dynamics formulated in terms of dumbbell vectors, on the one hand, and in terms of the conformation tensor, on the other hand. In this section, the relation of the dumbbell-vector description to the multiplicative decomposition of the conformation tensor (e.g., see \cite{HHA2018,HCHA2020}),
\bea
\cvh = \bvh_{\bnum^\prime} \cdot \bvh_{\bnum^\prime}^\Tr \, , 
\eea
 is examined, where $\bvh$ has dimensions $D \times \bnum^\prime$ with $\bnum^\prime$ an arbitrary dimension. This decomposition can be written in the form
\bea
\cvh = \sum_{\nu=1}^{\bnum^\prime} \bvh_{\bnum^\prime,\nu} \bvh_{\bnum^\prime,\nu} \, , 
\eea
with $\bvh_{\bnum^\prime,\nu}$ the $\nu$-th column vector of $\bvh_{\bnum^\prime}$ 
(see also \cite{publ72}). 
In general, one obviously must require $\bnum^\prime \ge D$ for this decomposition to be complete for arbitrary conformation tensor $\cvh$. 

In view of the expression \eqref{eq:ctensor} for the conformation tensor, a natural choice is $\bnum^\prime=N$ with $\bvh_{N,\mu} = \Qv_\mu/\sqrt{N}$, relating the dynamics of $\bvh_{N}$ directly to that of the dumbbell vectors $\Qv_\mu$. In the following, we focus on Hookean dumbbells, i.e., the Maxwell model, for illustrative purposes. The evolution equations \eqref{eq:generalSDE} for Hookean dumbbells, using the potential \eqref{eq:FENEP:phi} and mobility tensor \eqref{eq:Giesekus:M} with the parameters given in table~\ref{tab:models}, become
\bea
\dr\Qv_\mu = \left[\kv\cdot\Qv_\mu - \frac{1}{2 \lambda} \Qv_\mu \right] \dt
+ \frac{1}{\sqrt{\lambda}} \dr \Wv_\mu \, , \label{eq:LinkMaxwellRmu}
\eea
where the identifications $\lambda = \zeta/(4 H)$ and $H=\kBT$ have been made. Eq.~\eqref{eq:LinkMaxwellRmu} translates directly into the dynamics of $\bvh_{N}$,
\bea
\dr\bvh_N = \left[\kv\cdot\bvh_N - \frac{1}{2 \lambda} \bvh_N \right] \dt
+ \frac{1}{\sqrt{N \lambda}} \dr \check{\Wv}_N \, , \label{eq:LinkMaxwellBn}
\eea
where $\dr \check{\Wv}_N$ has dimensions $D \times N$ with its $\mu$-th column vector given by $\dr \Wv_\mu$. Let us now compare this result with the dynamics for the ``square root'' $\bv_D$ of the conformation tensor $\cv$, i.e. $\cv = \bv_D \cdot \bv_D^\Tr$ where $\bv_D$ has dimensions $D \times D$, as derived in \cite{HHA2018} and amended in \cite{HCHA2020},  
\bea
\dr \bv_D &=& \left[ \kv\cdot\bv_D - \frac{1}{2 \lambda}(\bv_D-\bv^{-1,\Tr}_D) - \frac{D}{2 N \lambda} \bv^{-1,\Tr}_D \right] \dr t \nonumber \\
{}&& + \frac{1}{\sqrt{N \lambda}} \dr \check{\Wv}_D \, . \label{eq:LinkMaxwellB3}
\eea
The close relation between $\bv_D$ and the elastic, i.e., recoverable, part of the deformation gradient in solid mechanics has been discussed in \cite{HHA2018}.
For $N \to \infty$, the dynamics of the column vectors of $\bv_3$ agree with the treatment proposed in \cite{Cho2009}.

Two major differences between eq.~\eqref{eq:LinkMaxwellBn} and eq.~\eqref{eq:LinkMaxwellB3} are apparent. First, the relaxation in eq.~\eqref{eq:LinkMaxwellBn} drives the column vectors of $\bvh_N$ to zero, while , considering e.g. $D=3$, according to eq.~\eqref{eq:LinkMaxwellB3} 
$\bv_{3,1}$ relaxes to $\bv_{3,2} \times \bv_{3,3} / \det{\bv_3}$,
$\bv_{3,2}$ to $\bv_{3,3} \times \bv_{3,1} / \det{\bv_3}$, and 
$\bv_{3,3}$ to $\bv_{3,1} \times \bv_{3,2} / \det{\bv_3}$,
respectively, i.e., the column vectors of $\bv_3$ become orthonormal in the course of relaxation. The second major difference between eq.~\eqref{eq:LinkMaxwellBn} and eq.~\eqref{eq:LinkMaxwellB3} relates to the absence of the thermal drift (the third term on the r.h.s. of eq.~\eqref{eq:LinkMaxwellB3}) in eq.~\eqref{eq:LinkMaxwellBn}. Both differences, in relaxation and thermal drift, are a hallmark of eliminating degrees of freedom when going to a reduced description of the dynamics. More specifically, both of these contributions are tightly related to the entropy, i.e., to the counting of states in configuration space, see sect.~\ref{sec:freeenergy}. It is pointed out that the difference in relaxation does not depend on the value of $N$, while the thermal drift clearly is a finite-$N$ effect, i.e., it is related to the fluctuations. However, despite these differences between  eq.~\eqref{eq:LinkMaxwellBn} and eq.~\eqref{eq:LinkMaxwellB3}, one should keep in mind that they both result in the same dynamics for the conformation tensor.

\subsection{Rotational dynamics}

In \cite{HCHA2020}, it has been discussed that, due to the non-unique\-ness of the decomposition $\cv = \bv_D \cdot \bv_D^\Tr$, there is on-going rotational dynamics in eq.~\eqref{eq:LinkMaxwellB3} even at equilibrium, with a relaxation time that depends approximately linearly on $N$. In the following, an attempt is made to rationalize this $N$-dependence of the rotational relaxation time in terms of the dynamics for the dumbbell vectors $\Qv_\mu$, eq.~\eqref{eq:LinkMaxwellRmu}. Consider the linear combination
\bea
\Rv = \sum^N_{\mu=1} \gamma_\mu \Qv_\mu \, . \label{eq:linearcomb}
\eea
Based on eq.~\eqref{eq:LinkMaxwellRmu}, its dynamics is given by
%%%%
\bea
\dr \Rv = \left[\kv\cdot\Rv - \frac{1}{2 \lambda} \Rv\right] \dt
+ \frac{1}{\sqrt{\lambda}} \dr \bar{\Wv} \, , \label{eq:generalSDEavg}
\eea
where the Wiener process increments defined by $\dr\bar{\Wv} = \sum^N_{\mu=1} \gamma_\mu \dr\Wv_\mu$ satisfy
\bea
\langle \dr \bar{\Wv} \rangle &=& \nullv \, , \\
\langle \dr \bar{\Wv} \dr \bar{\Wv} \rangle &=& \overline{\gamma^2} \dt \onev \, ,
\eea
with 
$\overline{\gamma^2} \equiv \sum^N_{\mu=1} \gamma_\mu^2 $.
Let us consider equilibrium, $\kv = \nullv$. In order to study the orientation dynamics, we write $\Rv = R \nv$, with $R$ and $\nv$ the length and the orientation vector of $\Rv$, respectively. By using It{\^o} calculus \cite{gardinerbook,hcostochbook}, it can be shown that the SDEs for  $R$ and $\nv$ are given by
\bea
\dr R &=& \left(\frac{\overline{\gamma^2}}{2 R \lambda} \left(D - 1\right) - \frac{1}{2 \lambda} R \right) \dt + \frac{1}{\sqrt{\lambda}} \nv \cdot \dr \bar{\Wv} \, , \\
\dr \nv &=& -\frac{\overline{\gamma^2}}{2 R^2 \lambda} \left(D - 1\right) \nv \dt + \frac{1}{R \sqrt{\lambda}} \left( \onev - \nv \nv \right) \cdot \dr \bar{\Wv} \, . \label{eq:rotdyn}
\eea
The contributions proportional to $\overline{\gamma^2}$ originate from the second-order term in the It{\^o} calculus \footnote{Note: It can be shown, again by using It{\^o} calculus \cite{gardinerbook,hcostochbook}, that indeed $\dr | \nv | = 0$.}. In particular, one observes in the orientation dynamics \eqref{eq:rotdyn} that the relaxation time for the orientation vector $\nv$ is given by
\bea
\lambda_n &=& \frac{2 R^2}{\overline{\gamma^2} \left(D - 1\right)} \lambda  \, . \label{eq:relaxtime}
\eea
As an example, consider the case $D = 3$. To get an idea about the rotational dynamics of the column vectors $\bv_{3,\mu}$ of $\bv_3$, which satisfy
$\cv = \sum_{\mu = 1}^3 \bv_{3,\mu} \bv_{3,\mu}$,
we make a particular choice for $\Rv$, i.e., for the linear combination \eqref{eq:linearcomb}. To that end, think of a Voronoi tessellation on the sphere, generated by the six ``poles'' that themselves are generated as intersections of three orthogonal axes with the sphere. Let us now consider two of these Voronoi sectors, $\mathcal{V}_{i}$ and $\mathcal{V}_{\bar{i}}$, which are on opposite sides of the sphere. It is reasonable to assume that the number of vectors $\Qv_\mu$ which have their orientation in these two Voronoi sectors together can be written as $\varphi N$, where $\varphi$ is independent of $N$. After mapping all vectors $\Qv_\mu$ with orientation in $\mathcal{V}_{\bar{i}}$ into $\mathcal{V}_{i}$ by way of $\Qv_\mu \to -\Qv_\mu$, which is just making use of the symmetry of dumbbell description, all $\varphi N$ vectors with orientation in $\mathcal{V}_{i}$ are averaged to obtain $\Rv$, i.e., $\gamma_\mu = 1/(\varphi N)$ for these vectors and $\gamma_\mu = 0 $ for all others. In this case, one finds 
$\overline{\gamma^2} = 1 / (\varphi N)$, and thus the relaxation time for rotation \eqref{eq:relaxtime} is increased w.r.t. $\lambda$, namely as 
$\lambda_n \propto N \lambda$, in agreement with the observation in \cite{HCHA2020}. 

For the construction of a suitable vector $\Rv$, the Voronoi sectors have been used, instead of, e.g., considering a random selection of vectors $\Qv_\mu$, for two reasons: First, the construction with Voronoi sectors allows to construct three such vectors (representative of the column vectors of $\bv_3$) that are clearly linearly independent from each other. And second, the length of $\Rv$ is proportional to the length of the dumbbell vectors $\Qv_\mu$ with a prefactor that is of order $\mathcal{O}(1/2)$, i.e. independent of $N$. 

\section{Discussion and conclusions} \label{sec:discussion}

The focus of this paper has been on deriving viscoelasticity in terms of a conformation tensor with fluctuations, based on the kinetic theory of dumbbells. This has been achieved by identifying the conformation tensor with the arithmetic average over a finite number $N$ of dumbbells, eq.~\eqref{eq:ctensor}, and using two alternative routes for deriving the dynamics: a direct approach using stochastic calculus, and a thermodynamic approach, in which the Helmholtz free energy plays a key role. It has been shown that these two approaches agree only if a finite-size contribution to the Helmholtz free energy of the conformation tensor is taken into account. \Rthree{The main messages of this paper are therefore the following: 
\begin{itemize}
\item If the number $N$ of dumbbells is {\it finite}, the commonly employed expressions for the thermodynamic potentials need to be corrected: Using statistical mechanics (see Appendix~\ref{sec:appendix1}), one finds that the conformational entropy must be corrected by replacing $N$ by $N-D-1$ (with $D$ the number of spatial dimensions), which in turn modifes the Helmholtz free energy, see eq.~\eqref{eq:psi} with correction term eq.~\eqref{eq:deltapsi}.
\item The thereby obtained finite-size correction in the free energy is crucial for guaranteeing compatibility between the dynamics of the conformation tensor and of the underlying dumbbells. 
\end{itemize}
While these general conclusions have been established in general terms for a large class of models, they have also been exemplified for the Hookean (Maxwell) model, the FENE-P model, and the Giesekus model (see Appendix~\ref{sec:appendix0}); the dynamics for the conformation tensor with fluctuations for these three models is summarized in table~\ref{table:finaldynamicmodels}.}

\begin{table*}[t]
\color{cref3}\centering
\caption{Dynamics for the conformation tensor $\cv$ with fluctuations; see Appendix~\ref{sec:appendix0} for the derivation. Symbols are explained in the text. Note that the square root of a tensor is symmetric. The increments $\dr\tilde{\Wv}$ of Wiener processes satisfy 
$\langle \dr\tilde{W}_{ij}(t) \rangle = 0$ and $\langle \dr\tilde{W}_{ij}(t) \, \dr\tilde{W}_{kl}(t^\prime) \rangle = \delta_{ik} \, \delta_{jl} \, \delta(t - t^\prime) \, dt \, dt^\prime$ (see eq.~\eqref{eq:noiseavgp65} and eq.~\eqref{eq:noisecorrp65}).
}
%the statistical properties specified by eq.~\eqref{eq:noiseavgp65} and eq.~\eqref{eq:noisecorrp65}.}
\label{table:finaldynamicmodels}
\vspace{0.5cm}
\begin{tabular}{lll}
\hline\hline
Model & Conformation dynamics & Noise \\\hline\\[-8truept]
{\bf Maxwell} &
$\displaystyle\dr \cv = \left[ \kv\cdot\cv + \cv\cdot\kv^\Tr 
- \frac{4 H}{\zeta} \left(\cv - \frac{\kBT}{H} \onev \right) \right] \dt 
+ \dr \cv^\fr $
&  $\displaystyle\dr \cv^\fr 
= \sqrt{\frac{4 \kBT}{N \zeta}} 
\left( \sqrt{\cv} \cdot \dr\tilde{\Wv}
+ \dr\tilde{\Wv}^\Tr \cdot \sqrt{\cv}
\right)$\\[8truept]
{\bf FENE-P}  & 
 $\displaystyle \dr \cv = \left[ \kv\cdot\cv + \cv\cdot\kv^\Tr 
- \frac{4 H}{\zeta} \left(f \, \cv - \frac{\kBT}{H} \onev \right) \right] \dt 
+ \dr \cv^\fr $ &
 $\displaystyle \dr \cv^\fr 
=  \sqrt{\frac{4 \kBT}{N \zeta}} 
\left(
\sqrt{\cv} \cdot \dr\tilde{\Wv}
+
\dr\tilde{\Wv}^\Tr \cdot \sqrt{\cv}
\right)
$ \\[10truept]
{\bf Giesekus} &
$\displaystyle \dr \cv = \left[ \kv\cdot\cv + \cv\cdot\kv^\Tr 
- \frac{4 H}{\zeta} \left((1-\alpha) \onev + \alpha \cv \right) \cdot \left(\cv - \frac{\kBT}{H} \onev \right) \right] \dt$ &
$\displaystyle \displaystyle \displaystyle\dr\cv^\fr =
\sqrt{\frac{4 \kBT}{N \zeta}} 
\left( \sqrt{\cv} \cdot \dr\tilde{\Wv} \cdot \bar{\Bv}
+ \bar{\Bv} \cdot \dr\tilde{\Wv}^\Tr \cdot \sqrt{\cv}
\right)
$ \\[10truept]
&$\qquad \displaystyle+\frac{4 \alpha (D+1) \kBT}{N \zeta} \cv \dt + \dr \cv^\fr $
 & \qquad where $\displaystyle\bar{\Bv} = \sqrt{(1-\alpha) \onev + \alpha \cv}$ \\[6truept]
\hline\hline
\end{tabular}
\vspace{1.0cm}
\color{black}
\end{table*}

When discussing a model with fluctuations, the deterministic counterpart serves as a benchmark. For all models discussed in this paper, one recovers the known deterministic models in the thermodynamic limit $N \to \infty$, i.e., if the number of dumbbells in the volume of interest $V$ diverges, keeping the number density $n = N/V$ constant. Beyond that thermodynamic limit, however, there is also an interest in the behavior of the average conformation tensor $\langle \cvh \rangle$ for finite $N$, i.e., in the presence of fluctuations. For most models studied in this paper, the nonlinearities do not allow to obtain a closed form equation for $\langle \cvh \rangle$ based on the stochastic differential equation (SDE), eq.~\eqref{eq:SDE:c:general} and eq.~\eqref{eq:SDEp65}, for the fluctuating conformation tensor $\cvh$. The notable exception to this rule is the Hookean dumbbell (Maxwell) model, with potential energy $\Phi_{\rm d}(\tr{\Qv \Qv}) = (H/2) \tr{\Qv \Qv}$ and mobility tensor $\Mv = (2/\zeta) \onev$. Taking the average of the SDE \eqref{APPeq:SDE:c:Maxwell} over different realizations of the fluctuations, one observes that the average conformation tensor $\langle \cvh \rangle$ obeys the same differential equation as its deterministic ($N \to \infty$) counterpart  --  an observation that generally does not hold for nonlinear models. In particular, one finds for the Hookean dumbbell model at equilibrium $\langle \cvh \rangle_{\textrm{eq}} = \onev$, where $H = \kBT$ has been used. As a word of caution, it is pointed out that the conformation tensor $\cvh$ that minimizes the Helmholtz free energy density, eq.~\eqref{eq:psi} with eq.~\eqref{eq:deltapsi}, is given by $\cvh_{\textrm{min}} = (1-(D+1)/N)\onev$, i.e., it does depend on the finite size ($N$) of the system. The fact that $\cvh_{\textrm{min}} \neq \langle \cvh \rangle_{\textrm{eq}}$ is not a contradiction; it merely points out that the distribution of thermal fluctuations around the minimum is not symmetric. 

The relevance of the finite-size correction of the Helm\-holtz free energy, $\Delta \psi$ given by eq.~\eqref{eq:deltapsi}, has been discussed primarily in the context of formulating dynamics with fluctuations for the conformation tensor $\cv$. However, it is also of immediate consequence for the formulation of fluctuating viscoelasticity in terms of the ``square root'' $\bv_3$, where $\cv = \bv_3 \cdot \bv_3^\Tr$. In \cite{HHA2018}, a thermodynamic approach has been taken towards deriving the dynamics of $\bv_3$, based on the dynamics of $\cv$. Therefore, if the thermodynamic potential for the $\cv$-dynamics contains a finite-size contribution, the same holds true also for the thermodynamic potential for the $\bv_3$-dynamics,  see sect.~4.3 in \cite{HHA2018} for details. Beyond these implications for the thermodynamics of a $\bv_3$-formulation, the kinetic models for $N$ dumbbells have also been employed in this paper to give an explanation for the existence of a rotational relaxation time proportional to $N$ in the fluctuating dynamics of $\bv_3$, which has been observed earlier \cite{HCHA2020}.

%%%%
\begin{acknowledgement}
MH acknowledges stimulating discussions with Hans Christian {\"O}ttinger, particularly in relation to the interpretation of eq.~\eqref{eq:LinkMaxwellB3}. PDO is grateful to Georgetown University and the Ives Foundation for support. 
A preliminary version of the derivation in Appendix~\ref{sec:app:geometry}, for the $N \to \infty$ limit, was developed by DJR during previous discussions with Joseph Peterson and Gary Leal.
\end{acknowledgement}

\section*{Author contribution statement}

Peter Olmsted initiated this collaboration of the authors, and Daniel Read spotted the inconsistency between the results in \cite{HHA2018} and a direct kinetic-theory approach.
Markus H\"utter has made the main contributions to the main part of the manuscript, as well as to Appendices~\ref{sec:appendix0}, \ref{sec:app:diffeqn} and \ref{sec:app:scaling:real}, while Daniel Read and Peter Olmsted have developed Appendices~\ref{sec:app:geometry} and \ref{sec:app:scaling:reciprocal}, respectively. All authors have discussed in detail about all parts of the paper, and helped to bring the paper in its final form. All the authors have read and approved the final manuscript.

\section*{Conflict of interest}

The authors declare that they have no conflict of interest.

\appendix
\numberwithin{equation}{section}

\section{Dynamics of conformation tensor for three exemplary models} \label{sec:appendix0}

For completeness, three concrete realizations of the general dynamics for the conformation tensor, eq.~\eqref{eq:SDEp65} with eq.~\eqref{eq:noisep65}, are provided in this section.

\subsection{Hookean, i.e., Maxwell, model} 
Using the potential $\Phi$ in eq.~\eqref{eq:FENEP:phi} with eq.~\eqref{eq:FENEP:f} in the limit $\beta_1 = 1$ and $\beta_2 \to \infty$, and with the mobility tensor $\Mv$ given by eq.~\eqref{eq:Giesekus:M} for $\alpha = 0$ for Hookean dumbbells, the general equation \eqref{eq:SDEp65} turns into the Maxwell model with fluctuations (see also \cite{HHA2018,HCHA2020}),
\bea
\dr \cvh &=& \left[ \kv\cdot\cvh + \cvh\cdot\kv^\Tr 
- \frac{4 H}{\zeta} \left(\cvh - \frac{\kBT}{H} \onev \right) \right] \dt 
%\nonumber \\ {}&&
+ \dr \cvh^\fr \, , \label{APPeq:SDE:c:Maxwell}
\eea
where the fluctuations are determined by eq.~\eqref{eq:noisep65} with eq.~\eqref{eq:decomposition}. In particular, one may choose both $\bv$ and $\Bv$ to have dimensions $D \times D$ and to be symmetric, i.e., $\bv = \sqrt{\cv}$ and $\Bv = \sqrt{\Mv} = \sqrt{2/\zeta} \, \onev$. This result for the stochastic dynamics of the conformation tensor is identical to what has been derived in \cite{vazquezquesadaetal2009}.

\subsection{FENE-P model} 
Using the potential $\Phi$ in eq.~\eqref{eq:FENEP:phi} with eq.~\eqref{eq:FENEP:f}, and with the mobility tensor $\Mv$ given by eq.~\eqref{eq:Giesekus:M} for $\alpha = 0$, the general equation \eqref{eq:SDEp65} turns into the FENE-P model with fluctuations (see also \cite{HHA2018,HCHA2020}),
\bea
\dr \cvh &=& \left[ \kv\cdot\cvh + \cvh\cdot\kv^\Tr 
- \frac{4 H}{\zeta} \left(f \cvh - \frac{\kBT}{H} \onev \right) \right] \dt 
%\nonumber \\ {}&&
+ \dr \cvh^\fr \, . \label{APPeq:SDE:c:FENEP}
\eea
The fluctuations are determined by eq.~\eqref{eq:noisep65} and eq.~\eqref{eq:decomposition} where, as for the Maxwell model, one may choose both $\bv$ and $\Bv$ to have dimensions $D \times D$ and to be symmetric, i.e., $\bv = \sqrt{\cv}$ and $\Bv = \sqrt{\Mv} = \sqrt{2/\zeta} \, \onev$.

\subsection{Giesekus model} 

Using the potential $\Phi$ in eq.~\eqref{eq:FENEP:phi} with eq.~\eqref{eq:FENEP:f} in the limit $\beta_1 = 1$ and $\beta_2 \to \infty$, and with the mobility tensor $\Mv$ given by eq.~\eqref{eq:Giesekus:M}, the general equation \eqref{eq:SDEp65} turns into the Giesekus model with fluctuations (see also \cite{HHA2018,HCHA2020}),
\bea
\dr \cvh &=& \left[ \kv\cdot\cvh + \cvh\cdot\kv^\Tr 
- \frac{4 H}{\zeta} \left((1-\alpha) \onev + \alpha \cvh \right) \cdot \left(\cvh - \frac{\kBT}{H} \onev \right) \right] \dt \nonumber \\
{}&&
+\frac{4 \alpha (D+1) \kBT}{N \zeta} \cvh \dt + \dr \cvh^\fr \, , \label{APPeq:SDE:c:Giesekus}
\eea
where the fluctuations are determined by eq.~\eqref{eq:noisep65} with eq.~\eqref{eq:decomposition}. One may choose both $\bv$ and $\Bv$ to have dimensions $D \times D$ and to be symmetric, i.e., $\bv = \sqrt{\cv}$ and $\Bv = \sqrt{\Mv}$, where $\Mv$ is given by eq.~\eqref{eq:Giesekus:M} for $\alpha \neq 0$. The thereby-obtained expression for the fluctuations differs from that used in \cite{HHA2018}, while sharing the same statistical properties. The difference in the expressions for the fluctuations reflects the non-uniqueness of the decomposition \eqref{eq:p65:FDT}.

Of the three models discussed explicitly in this Appendix \ref{sec:appendix0}, the Giesekus model is the only one for which there is a thermal drift in the dynamics of $\cvh$, i.e., the second-last contribution on the r.h.s. of eq.~\eqref{APPeq:SDE:c:Giesekus}. It is present only if the mobility of the dumbbells is anisotropic, $\alpha \neq 0$. Beyond the $N$-dependence of the noise $\dr \cvh^\fr$, the thermal drift is the second explicit consequence of the finite size of the system.

\section{Calculation of partition function $\Gamma$} \label{sec:appendix1}

In this Appendix, three procedures are presented for calculating the microcanonical partition function $\Gamma$, eq.~\eqref{eq:Gamma:general}, for a finite number $N$ of dumbbells.

\subsection{Procedure 1: Differential equation} \label{sec:app:diffeqn}

We start by noting that the microcanonical partition function $\Gamma$,  eq.~\eqref{eq:Gamma:general}, contains only $K = D(D+1)/2$ (rather than $D^2$) Dirac $\delta$-functions, representative
of the constraints related to the independent components
$c_{ij}$ with $1 \le i \le j \le D$. The following notation is introduced: 
\bea
\Delta_{ij} &=& \delta\left( \ch_{ij}-c_{ij}\right) \, , \qquad i \le j \, , \label{eq:app:constraint}\\
\Delta^{[K]} &=& \prod_{k,l=1; k \le l}^D \Delta_{kl} \, , \label{eq:app:DeltaK}\\
\Delta^{[K-1]}_{ij} &=& \prod_{k,l=1; k \le l; 
%(k,l) \neq (i,j)
k\neq i; l\neq j
}^D \Delta_{kl} \, , \qquad i \le j \, , \label{eq:app:DeltaKmone}
\eea
so that $\Delta^{[K-1]}_{ij}\Delta_{ij}=\Delta^{[K]}$. With eq.~\eqref{eq:app:DeltaK}, the microcanonical partition function can be written in the form
\bea
\Gamma = \int \Delta^{[K]} \, d^{DN}Q \, . \label{eq:app:Gammarewritten}
\eea
One can show that
\bea
\frac{\partial \Gamma}{\partial c_{ij}} = - \int \Delta^{[K-1]}_{ij} \Delta^\prime_{ij}\, d^{DN}Q \, , \qquad i \le j \, , 
\label{eq:app:Gammader}
\eea
where $\Delta^\prime_{ij}$ denotes the derivative of $\Delta_{ij}$ w.r.t. its argument. Furthermore, one can derive the following relations, where  $Q_{\mu,k}$ is the $k^{\textrm{th}}$ component of the $\mu^{\textrm{th}}$ dumbbell,
\bea
\sum_{\mu=1}^N Q_{\mu,k} \frac{\partial \Delta_{ij}}{\partial Q_{\mu,k}} 
= 
\left( \delta_{ik} \ch_{jk} + \delta_{jk} \ch_{ik} \right)
\Delta^\prime_{ij} \, , \qquad i \le j \, , \label{eq:identity}
\eea
of which there are three non-zero  cases: 
%\end{document}
\bea
i = j = k : && \sum_{\mu=1}^N Q_{\mu,i} \frac{\partial \Delta_{ii}}{\partial Q_{\mu,i}} 
= 2 \left(\Delta^\prime_{ii} c_{ii} - \Delta_{ii}\right) \, ,  \label{eq:app:case1} \\
i  < j, k = i : && \sum_{\mu=1}^N Q_{\mu,i} \frac{\partial \Delta_{ij}}{\partial Q_{\mu,i}} 
= \Delta^\prime_{ij} c_{ij} - \Delta_{ij} \, ,  \label{eq:app:case2} \\
i < j, k = j : && \sum_{\mu=1}^N Q_{\mu,j} \frac{\partial \Delta_{ij}}{\partial Q_{\mu,j}} 
= \Delta^\prime_{ij} c_{ij} - \Delta_{ij} \, .  \label{eq:app:case3}
\eea
In calculating eqs.~\eqref{eq:app:case1}--\eqref{eq:app:case3}, we have made use of the identity
$x \delta^\prime(x) = -\delta(x)$, 
with $x = \ch_{ij}-c_{ij}$, which implies
$(\ch_{ij}-c_{ij}) \delta^\prime(\ch_{ij}-c_{ij}) = -\delta(\ch_{ij}-c_{ij})$, from which it follows $\ch_{ij} \Delta^\prime_{ij} = c_{ij} \Delta^\prime_{ij} -\Delta_{ij}$.
To proceed, we use eq.~\eqref{eq:app:Gammader} for $i = j$ and eq.~\eqref{eq:app:case1}  to derive an expression for $2 c_{ii} (\partial \Gamma / \partial c_{ii})$, 
\bea
2 c_{ii} \frac{\partial \Gamma}{\partial c_{ii}} &=&-
\int \Delta^{[K-1]}_{ii} 
\left[\sum_{\mu=1}^N Q_{\mu,i} \frac{\partial \Delta_{ii}}{\partial Q_{\mu,i}} +2\Delta_{ii}\right]\, d^{DN}Q \nonumber \\
{} &=& -2 \Gamma -\int \Delta^{[K-1]}_{ii} 
\left[\sum_{\mu=1}^N Q_{\mu,i} \frac{\partial \Delta_{ii}}{\partial Q_{\mu,i}}\right]\, d^{DN}Q \, , \nonumber \\
\eea
where we have used $\Delta^{[K-1]}_{ii}\Delta_{ii}=\Delta^{[K]}$. The remaining integral is then re-written by performing an integration by parts w.r.t. $Q_{\mu,i}$; the corresponding boundary-terms can be neglected if the values of the components of $\cv$ are finite. This leads to
\bea
2 c_{ii} \frac{\partial \Gamma}{\partial c_{ii}} 
&=& (N-2) \Gamma +\int \Delta_{ii} \sum_{\mu=1}^N Q_{\mu,i} 
\frac{\partial \Delta^{[K-1]}_{ii}}{\partial Q_{\mu,i}} \, d^{DN}Q \, , 
\nonumber \\
\eea
where, again, we have used  $\Delta^{[K-1]}_{ii}\Delta_{ii}=\Delta^{[K]}$.
Using the product rule for calculating the derivative $\partial \Delta^{[K-1]}_{ii}/\partial Q_{{\mu,i}}$ and rearranging terms results in
\bea
2 c_{ii} \frac{\partial \Gamma}{\partial c_{ii}} = (N-2) \Gamma 
+ \sum_{j=1}^{i-1} G_{ji} + \sum_{j=i+1}^{D} G^\star_{ij} \, , \label{eq:app:intermed1}
\eea
where the quantities $G_{ji}$ and $G^\star_{ij}$are given by 
\bea
G_{ji} &\equiv& \int \Delta^{[K-1]}_{ji} 
\sum_{\mu=1}^N Q_{\mu,i} \frac{\partial \Delta_{ji}}{\partial Q_{\mu,i}} \, d^{DN}Q \nonumber \\
{}&=& - c_{ji} \frac{\partial \Gamma}{\partial c_{ji}} - \Gamma \, , \qquad j < i \, , \label{eq:app:intermed3} \\  
G^\star_{ij} &\equiv& \int \Delta^{[K-1]}_{ij} 
\sum_{\mu=1}^N Q_{\mu,i} \frac{\partial \Delta_{ij}}{\partial Q_{\mu,i}} \, d^{DN}Q \nonumber \\
{}&=& - c_{ij} \frac{\partial \Gamma}{\partial c_{ij}} - \Gamma \, , \qquad i < j\, , \label{eq:app:intermed2} 
\eea
where we have used eq.~\eqref{eq:app:case2} and eq.~\eqref{eq:app:case3} (with $i$ and $j$ interchanged), and then eq.~\eqref{eq:app:Gammarewritten} and eq.~\eqref{eq:app:Gammader}. Combining eq.~\eqref{eq:app:intermed1} with eq.~\eqref{eq:app:intermed2} and eq.~\eqref{eq:app:intermed3}, one obtains
\bea
\sum_{j=1}^{i-1} c_{ji} \frac{\partial \Gamma}{\partial c_{ji}}  
+ 2 c_{ii} \frac{\partial \Gamma}{\partial c_{ii}} 
+ \sum_{j=i+1}^{D} c_{ij} \frac{\partial \Gamma}{\partial c_{ij}}
= (N-D-1) \Gamma \, , \nonumber \\ {}&& \label{eq:app:diffeqn}
\eea
which is a differential equation for $\Gamma$. To solve this equation, consider the ansatz
\bea
\Gamma = \Gamma_0 \left( \det\cv \right)^\nu \, . \label{eq:app:ansatz1}
\eea
For calculating the partial derivatives of this ansatz w.r.t. $c_{ij}$, the following needs to be kept in mind. On the one hand, since $\cv$ is symmetric, there are only $K$ independent variables, rather than $D^2$, which is in line with the strategy adopted, e.g., in eq.~\eqref{eq:app:DeltaK} and eq.~\eqref{eq:app:DeltaKmone}. This implies that there are only $K$ partial derivatives to be calculated in eq.~\eqref{eq:app:diffeqn}, namely w.r.t. $c_{ij}$ with $1 \le i \le j \le D$. On the other hand,  the tensor $\cv$ contains elements $c_{ij}$ and $c_{ji}=c_{ij}$. 
Hence, for $i \neq j$, one finds for any function $f(\cv)$: $\partial f(\cv)/\partial c_{ij} = [\partial f(\cv)/\partial c_{ij} ]_{\rm no-sym} + [\partial f(\cv)/\partial c_{ji} ]_{\rm no-sym}$, where ``no-sym'' emphasizes that the corresponding derivative is taken without enforcing the symmetry of $\cv$, i.e., considering all $D^2$ components of $\cv$ as independent variables. For the derivative of ansatz \eqref{eq:app:ansatz1} one thus  obtains
\bea
\frac{\partial \Gamma}{\partial c_{kl}} = (2-\delta_{kl}) \nu \Gamma \left(\cv^{-1}\right)_{kl} \, , \qquad k \le l \, , 
\label{eq:app:ansatz2}
\eea
where the factor $(2-\delta_{kl})$ originates from the fact that there are only $K$ independent components in $\cv$, $c_{kl}$ with $k \le l$; particularly, for $k < l$, the variable $c_{kl}$ appears at two off-diagonal positions in the full, symmetric matrix $\cv$. With this, it can be shown that the ansatz \eqref{eq:app:ansatz1} is indeed a solution of the differential equation \eqref{eq:app:diffeqn} if 
\bea
\nu = \frac{N-D-1}{2} \, .
\eea
The $\cv$-independent prefactor $\Gamma_0$ can not be determined with this approach. However, since $\Gamma_0$ results only in an additive contribution to the Helmholtz free energy $\Psi$, it turns out to be irrelevant for the dynamics of the conformation tensor $\cv$.

\subsection{Procedure 2: Geometry} \label{sec:app:geometry}

We begin with a geometrical interpretation of the microcanonical partition-function $\Gamma(\cv)$ defined via eqs.~\eqref{eq:Z:factors} and \eqref{eq:Gamma:general}.  Substituting from eq.~\eqref{eq:Z:factors} back into the full partition function eq.~\eqref{eq:Xi:full} we find:  
\bea
\Xi &=& \int e^{-\Phi(\cv)/(\kBT)} \, \Gamma(\cv) \, d^K c \nonumber \\
  &=& \int e^{-\Phi(\cvh)/(\kBT)} \, d^{D N}Q . \label{eq:Zfull1}
\eea
 Comparing the two lines of this equation we recognise that we may interpret the differential quantity $\Gamma(\cv) \, d^K c$ as being the volume, within the $DN$ dimensional $\{\Qv_\mu\}$-space, that is within an increment $d^K c$ of $\cv$ (i.e. where the $K$ independent components of $\cv$ are each varied within an interval $dc_\alpha$ of their base value). Evaluation of the dependence of this volume on $\cv$ yields $\Gamma(\cv)$ up to a constant prefactor.  Since this derivation requires some visual imagination, we first demonstrate how this may be reasoned in the specific case $D=3$ before indicating how the calculation may be generalised to arbitrary $D$.
\subsubsection{Three dimensions}
For $D=3$, $\{\Qv_\mu\}$-space is a $3N$ dimensional space spanned by the vectors $\{\Qv_\mu\}_{\mu=1,\ldots,N}$. For convenience, we specify a location within this space via the $3N$ dimensional vector $(\XX, \YY, \ZZ)$, where $\XX = (x_1,x_2,x_3 \ldots ,x_N)$ is the $N$-dimensional vector of the $x$-components of $\{\Qv_\mu\}_{\mu=1,\ldots,N}$ (and similarly for $\YY$ and $\ZZ$). 

We also note that, for $D=3$, there are $K=6$ independent components of the symmetric $\cv$ tensor: $c_{xx}$, $c_{yy}$, $c_{zz}$, $c_{xy}$, $c_{xz}$ and $c_{yz}$.   

We proceed by evaluating the dependence on $\cv$ of two separate volumes. We first evaluate $\Vconst$, which is the $(3N-6)$-dimensional subvolume of $\{\Qv_\mu\}$-space within which $\cvh$ (given in terms of $\{\Qv_\mu\}$ by eq.~\ref{eq:ctensor}) is held exactly equal to $\cv$.  Then, for each point within $\Vconst$, we find the volume $dV_6$ which is the $6$-dimensional subvolume swept out by varying $c_{xx}$ by $dc_{xx}$, $c_{yy}$ by $dc_{yy}$, etc.  Such excursions must all be perpendicular to the subvolume $\Vconst$ (in the $\{\Qv_\mu\}$-space), because contours of fixed $c_{xx}$ are perpendicular to the gradient direction of $c_{xx}$ (etc.). So, we can evaluate $\Gamma(\cv) \, d^6\cv$ as the product of $\Vconst$ and $dV_6$:
\[
\Gamma(\cv) \, d^6\cv = \Vconst \ dV_6 \, .
\]

Evaluation of these volumes is most straightforward in the co-ordinate frame in which 
$\cv$ is diagonalised (which can always be done since $\cv$ is symmetric).  
In the diagonal frame, $\cv$ takes values $\lambda_x$, $\lambda_y$ and $\lambda_z$ 
along the diagonal, but is zero in the off diagonal components. In terms of the $N$-dimensional vectors $\XX$, $\YY$ and $\ZZ$ these constraints can be expressed as:
\begin{align*}
    \left| \XX \right|^2 = \Lambda_x \mathrm{,~} \left| \YY \right|^2 &= \Lambda_y \mathrm{,~} \left| \ZZ \right|^2 = \Lambda_z, \\
    \XX \cdot \YY = 0 \mathrm{,~} \YY \cdot \ZZ &= 0 \mathrm{,~} \XX \cdot \ZZ = 0,
\end{align*}
where $\Lambda_i=N\lambda_i$, i.e. the vectors $\XX$, $\YY$ and $\ZZ$ are restricted to the surfaces of $N$-dimensional hyperspheres of radii $\sqrt{\Lambda_x}$, $\sqrt{\Lambda_y}$, and $\sqrt{\Lambda_z}$ respectively, whilst simultaneously being held to be mutually perpendicular. 

The volume $\Vconst$ is the $(3N-6)$-dimensional volume swept out by rotating the $\XX$, $\YY$ and $\ZZ$ vectors subject to the above constraints, the rotations being within the $N$-dimensional space of these vectors. A brief analogy may help at this stage: the surface of a sphere of radius $r$ is found by summing up small tiles formed by varying the polar co-ordinate angles $\theta$ and $\phi$ by small increments $d\theta$ and $d\phi$.  The distance along the sphere surface moved during such increments is $r d\theta$ and $r d\phi$ (multiplied by a geometric factor $\sin \theta$ which is irrelevant to the scaling with $r$). So, the scaling of surface area with $r$ can be found from the product of these lengths, $r^2 d\theta d\phi$. Likewise, the volume $\Vconst$ is the sum over a tiling of small incremental volumes, made by rotating the vectors $\XX$, $\YY$ and $\ZZ$ by small angles $d \theta_k$ in each available direction, whilst keeping their length fixed and retaining their mutually perpendicular orientation.  
% We consider first rotations of the $\XX$ vector. 
Since $\XX$ has $N$ dimensions, there are in total $(N-1)$ directions in which $\XX$ could be rotated whilst keeping the length of $\XX$ fixed. One such rotation will rotate the $\XX$ vector towards the $\YY$ vector. In this case, the $\YY$ vector must also rotate by the same angle, so as to maintain the perpendicular condition $\XX \cdot \YY = 0$. 
Hence, for rotation angle $d \theta_1$, $\XX$ rotates so that its end sweeps out a length $dl_X = \sqrt{\Lambda_x} d\theta_1$ perpendicular to $\XX$ (in the direction of $\YY$). Likewise $\YY$ also rotates so that its end sweeps out a length $dl_Y = \sqrt{\Lambda_y} d\theta_1$ perpendicular to $\YY$ (in the direction of $-\XX$). These changes in $\XX$ and $\YY$ result in a total length moved in $\{\Qv_\mu\}$-space which is 
\begin{align*}
    dl_1 &= \sqrt{dl_X^2 + dl_Y^2} \\
       &= \sqrt{\Lambda_x + \Lambda_y} d \theta_1.
\end{align*}
Similarly a second rotation direction of the $\XX$ vector is available, towards the $\ZZ$ vector (in which case $\ZZ$ must also rotate). Likewise, a third rotation carries $\YY$ towards $\ZZ$ without rotating $\XX$. By similar arguments, the total length moved in $\{\Qv_\mu\}$-space for these is:
\begin{eqnarray*}
    dl_2 &=& \sqrt{\Lambda_x + \Lambda_z} d \theta_2, \\
    dl_3 &=& \sqrt{\Lambda_y + \Lambda_z} d \theta_3.
\end{eqnarray*}
These three rotation directions having been dealt with, there remain $(N-3)$ further rotation directions available for the $\XX$ vector, all of which allow $\XX$ to remain perpendicular to both $\YY$ and $\ZZ$. For each of these rotations of $\XX$, the length moved in $\{\Qv_\mu\}$-space is
\begin{equation*}
    dl_k = \sqrt{\Lambda_x} d \theta_k, \, k=4 \ldots N.
\end{equation*}
Similarly, there are $(N-3)$ rotations available to each of $\YY$ and $\ZZ$, each of which leaves the other vectors unchanged, giving lengths:
\begin{eqnarray*}
    dl_k &=& \sqrt{\Lambda_y} d \theta_k, \, k=(N+1) \ldots (2N-3), \\
    dl_k &=& \sqrt{\Lambda_z} d \theta_k, \, k=(2N-2) \ldots (3N-6).
\end{eqnarray*}
%Having considered rotations of the $\XX$ vector we turn to the remaining available rotations of the the $\YY$ vector. We have already considered rotations of $\YY$ in the direction $\XX$.  There remains rotation of $\YY$ in the direction $\ZZ$ which sweeps out a length 
%\begin{equation*}
%
%\end{equation*}
%and $(N-3)$ further rotations available for the $\YY$ vector which allow $\YY$ to remain perpendicular to both $\XX$ and $\ZZ$, each of which sweeps out a length
%\begin{equation*}
%    dl_k = \sqrt{\lambda_y} d \theta_k, \, k=(N+1) \ldots (2N-3).
%\end{equation*}
%Finally for $\ZZ$ we have already considered rotations towards both $\XX$ and $\YY$. There remain $(N-3)$ further rotations available, each of which sweeps out a length 
%\begin{equation*}
%    dl_k = \sqrt{\lambda_z} d \theta_k, \, k=(2N-2) \ldots (3N-6).
%\end{equation*}
A single ``tile" in the volume $\Vconst$ is obtained by sweeping through each of the above $(3N-6)$ incremental rotations. The total volume $\Vconst$ is obtained by adding together all such tiles as the vectors are rotated. We require only the dependence of $\Vconst$ on $\cv$ (ignoring prefactors).  Hence, by taking the product of the lengths swept out by a set of incremental rotations, we obtain:
\[
\prod_{k=1}^{3N-6} dl_k \sim \Vconst \prod_{k=1}^{3N-6} d\theta_k
\]
where
\begin{align}
    \Vconst &\sim \left(\sqrt{\Lambda_x \Lambda_y \Lambda_z}\right)^{(N-3)} \sqrt{\Lambda_x + \Lambda_y} \sqrt{\Lambda_x + \Lambda_z}\sqrt{\Lambda_y + \Lambda_z}.
\end{align}

We now turn to the volume $dV_6$, the $6$-dimensional subvolume swept out by incrementing $c_{xx}$ by $dc_{xx}$, $c_{yy}$ by $dc_{yy}$, etc. away from a single point within $\Vconst$.  We first calculate the length in $\{\Qv_\mu\}$-space traversed by incrementing $c_{xx}$ by $dc_{xx}$ in a direction perpendicular to the contour of constant $c_{xx}$. We note that
\[
c_{xx} = \XX \cdot \XX/N,
\]
so the direction perpendicular to the contour of constant $c_{xx}$ is
\[
\nabla c_{xx} = (2\XX/N,\mathbf{0},\mathbf{0}) \, .
\]
Moving a distance $dl_{xx}$ in this direction changes $c_{xx}$ by
\begin{align*}
d c_{xx} &= \left| \nabla c_{xx} \right| dl_{xx}  \\
         &= (2 \left| \XX \right|/N) dl_{xx} \\
         &= (2 \sqrt{\Lambda_x}/N) dl_{xx}.
\end{align*}
Inverting this, the distance moved in $\{\Qv_\mu\}$-space when incrementing $c_{xx}$ by $dc_{xx}$ is
\[
dl_{xx} = \frac{N}{2\sqrt{\Lambda_x}} d c_{xx} 
\]
and similarly for $d c_{yy}$ and $d c_{zz}$.

We next note that
\[
c_{xy} = \XX \cdot \YY/N,
\]
so the direction perpendicular to the contour of constant $c_{xy}$ is
\[
\nabla c_{xy} = (\YY/N,\XX/N,\mathbf{0})
\]
Moving a distance $dl_{xy}$ in this direction changes $c_{xy}$ by
\begin{align*}
d c_{xy} &= \left| \nabla c_{xy} \right| dl_{xy}  \\
         &= \left( \sqrt{\left| \XX \right|^2 + \left| \YY \right|^2}/N \right) dl_{xy} \\
         &= (\sqrt{\Lambda_x + \Lambda_y} /N) dl_{xy}.
\end{align*}
Inverting this, the distance moved in $\{\Qv_\mu\}$-space when incrementing $c_{xy}$ by $dc_{xy}$ is
\[
dl_{xy} = \frac{N}{\sqrt{ \Lambda_x + \Lambda_y}} d c_{xy} 
\]
and similarly for $d c_{xz}$ and $d c_{yz}$.

Hence, multiplying these six lengths together,
\begin{equation}
    dV_6 \sim \frac{N^6 d^6\cv}{\sqrt{\Lambda_x \Lambda_y \Lambda_z} \sqrt{\Lambda_x + \Lambda_y} \sqrt{\Lambda_x +\Lambda_z}\sqrt{\Lambda_y + \Lambda_z}}. 
\end{equation}
Evaluating $\Gamma(\cv) \, d^6\cv  = \Vconst \ dV_6$, and hence $\Gamma(\cv)$, we find that all factors of form $\sqrt{ \Lambda_x + \Lambda_y}$ cancel, and we are left with
\begin{eqnarray}
    \Gamma(\cv) &\sim& N^6 \left( \Lambda_x \Lambda_y \Lambda_z \right)^{(N-4)/2}  \nonumber \\
    &\sim& N^{3N/2} \left( \lambda_x \lambda_y \lambda_z \right)^{(N-4)/2}.
\end{eqnarray}
But $\mathrm{det}\cv = \lambda_x \lambda_y \lambda_z$, and so: 
\begin{equation}
    \Gamma(\cv) \sim  N^{3N/2} \left( \mathrm{det}\cv \right)^{(N-4)/2}
\end{equation}
with dependence on $\mathrm{det}\cv$ as required.

\subsubsection{Generalisation to $D$ dimensions}

Most of the above formalism carries directly over to $D$ dimensions.  
In calculating $\Vconst$, each vector of type $\XX$ has a total number of $(N-1)$ rotations available (whilst preserving length), but $(D-1)$ of these rotate the vector towards one of the others.  So, the total number of rotations available for each vector, which do not also require rotating one of the other vectors, is $(N-D)$. The number of rotations requiring two vectors to rotate (i.e. of the form $\XX$ to $\YY$, $\YY$ to $\ZZ$ etc., but avoiding double counting) is \linebreak $M=D(D-1)/2$.   So, $\Vconst$ is of form:  
\begin{align*}
    \Vconst \sim  &\left( \Lambda_x \Lambda_y \Lambda_z \ldots \right)^{(N-D)/2} \\
                  & \times \sqrt{\Lambda_x + \Lambda_y} \sqrt{\Lambda_x +\Lambda_z}\sqrt{\Lambda_y + \Lambda_z} \ldots,
\end{align*}
where there are $M$ terms of form $\sqrt{\Lambda_x + \Lambda_y}$. 

In calculating $dV_K$, there are $D$ diagonal constraints and $M=D(D-1)/2$ off-diagonal constraints in $\cv$ (a total of $K=D(D+1)/2$). Each diagonal constraint gives a factor of form $\sqrt{\Lambda_x}$ in the denominator of $dV_K$, whilst each off diagonal constraint gives a factor of form $\sqrt{\Lambda_x + \Lambda_y}$. Hence:
\[
dV_K \sim \frac{N^K d^K\cv}{ \sqrt{\Lambda_x \Lambda_y \Lambda_z \ldots } \sqrt{\Lambda_x + \Lambda_y} \sqrt{\Lambda_x +\Lambda_z}\sqrt{\Lambda_y + \Lambda_z} \ldots}.
\]
In the product $\Gamma(\cv) \, d^K\cv = \Vconst \ dV_{K}$, all $M$ factors of form $\sqrt{\Lambda_x + \Lambda_y}$ are present once in the numerator, and once in the denominator, and so cancel.
Hence:
\begin{align}
\Gamma(\cv) \, d^K\cv &= \Vconst . dV_K \nonumber \\
  &\sim N^K \left( \Lambda_x \Lambda_y \Lambda_z \ldots \right)^{(N-D-1)/2}d^K\cv \nonumber \\
  &\sim N^{DN/2} \left( \lambda_x \lambda_y \lambda_z \ldots \right)^{(N-D-1)/2}d^K\cv \nonumber \\
  &\sim  N^{DN/2} \left( \mathrm{det}\cv \right)^{(N-D-1)/2}d^K\cv .
\end{align}

\subsection{Procedure 3: Scaling}

In this procedure, scaling arguments are employed for the calculation of the microcanonical partition function $\Gamma$,  eq.~\eqref{eq:Gamma:general}, in real space and in reciprocal space, respectively.

\subsubsection{Scaling in real space} \label{sec:app:scaling:real}

Similar to the procedure described in sec.~\ref{sec:app:geometry}, it is again chosen to describe the $\{\Qv_\mu\}$-space in terms of the $N$-dimensional vectors $\XX_i$, $1 \le i \le D$, where the $\mu$-th component of $\XX_i$ equals the $i$-th component of $\Qv_\mu$. 

In what follows, we consider the coordinate system in which the conformation tensor $\cv$ is diagonal, with eigenvalues $\lambda_i$ $(1 \le i \le D)$. The microcanonical partition function $\Gamma$,  eq.~\eqref{eq:Gamma:general} with instantaneous conformation tensor $\cvh = \cvh(\Qv)$, can be written in the $\{\XX_i\}$-representation as 
\bea
\Gamma(\cv) = \int \, \delta^{(K)}\left(\cvh(\XX)-\cv\right) \, 
\prod_{i=1}^D \prod_{\mu=1}^N dX_{i,\mu} \, , \label{eq:Gamma:general:XX}
\eea
with $\cvh = \cvh(\XX)$ given by $\ch_{ij} = \XX_i \cdot \XX_j / N$. In view of the conditions $\ch_{ii} = c_{ii} = \lambda_i$, the integral \eqref{eq:Gamma:general:XX} is re-written by introducing $\XX_i = \sqrt{\lambda_i} \tilde{\XX}_i$. Making use of the property $\delta(a x) = (1/a) \delta(x)$ for $a > 0$, one obtains from eq.~\eqref{eq:Gamma:general:XX},
\bea
\Gamma(\cv) = \Gamma_1(\cv) \, \Gamma_2(\cv) \, \Gamma_3 \, , 
\eea
with
\bea
\Gamma_1(\cv) &=& \prod_{i=1}^D \prod_{\mu=1}^N \sqrt{\lambda_i} 
= (\det\cv)^{N/2} \, , \\
\Gamma_2(\cv) &=& \prod_{1 \le i \le j \le D} \frac{1}{\sqrt{\lambda_i \lambda_j}} \nonumber \\
{}&=& \sqrt{
\left(\prod_{1 \le i, j \le D} \frac{1}{\sqrt{\lambda_i \lambda_j}}\right)
\left(\prod_{1 \le i \le D} \frac{1}{\lambda_i}\right)
} \nonumber \\
{}&=& \sqrt{ (\det\cv)^{-D} (\det\cv)^{-1} } \nonumber \\
{}&=& (\det\cv)^{(-D-1)/2} \, , \\
\Gamma_3 &=& \int \, \delta^{(K)}\left(\cvh(\tilde{\XX})-\onev\right) \, 
\prod_{i=1}^D \prod_{\mu=1}^N d\tilde{X}_{i,\mu}\, ,
\eea
where $\Gamma_1$ comes from the substitution of variables in the volume element, and $\Gamma_2$ comes from the substitution of variables in $\delta^{(K)}$. It is to be noted that the integral $\Gamma_3$ does {\it not} depend on $\cv$. Rather, we find $\Gamma_3 = \Gamma(\onev)$. Therefore, one obtains for the $\cv$-dependence of the microcanonical partition function
\bea
\Gamma(\cv) = (\det\cv)^{(N-D-1)/2} \Gamma(\onev) \, .\label{eq:gamma1}
\eea

\subsubsection{Scaling in reciprocal space}  \label{sec:app:scaling:reciprocal}

An alternative derivation for $\Gamma(\cv)$ follows by  explicitly enforcing the constraints using Fourier transforms, and then performing a rescaling.
We begin with the microcanonical partition-function $\Gamma$ defined in eq.~\eqref{eq:Gamma:general},
\bea
\Gamma(\cv)&=&  \int d^{DN}Q\,\delta^{(K)}\left(\cv -\cvh\right) \, ,
%\noalign{\noindent\textrm{where}}
%\cvh&=\frac{1}{N}\sum_{\mu}\Qv_{\mu} \Qv_{\mu}\label{eq:chat}
\eea
where $\cvh$ is the symmetric number-averaged conformation tensor given by eq.~\eqref{eq:ctensor}. The $\delta$-function ensures that only configurations $\{\Qv_{\mu}\}$ that satisfy $\cv=\cvh$ are included in $\Gamma(\cv)$. To evaluate $\Gamma(\cv)$ we choose a coordinate system for the $\Qv_{\mu}$ comprising the  eigenvectors $\{\uvec{e}_a\}$ and eigenvalues $\lambda_a$ of $\cv$. The constraints in the Dirac $\delta$-function then reduce to 
\bea
    \lambda_a &=& \uvec{e}_a\cdot\cvh\cdot\uvec{e}_a \quad (a=1,\dots D) \quad D\,\textrm{constraints}, \\
     0 &=& \uvec{e}_b\cdot\cvh\cdot\uvec{e}_c \quad (b < c) \quad L\,\textrm{constraints},
\eea
where there are $L=D(D-1)/2$ zero off-diagonal elements. 
Hence we introduce $D$ Dirac $\delta$-functions for the diagonal constraints, and $L$ $\delta$-functions for the off-diagonal constraints. We enforce these with Fourier transforms, leading to
\bea
\Gamma(\cv) &=& \int \frac{d^D\!u}{(2\pi)^D}\int \frac{d^{L}\!v}{(2\pi)^L}\int d^{DN}Q \nonumber \\
&& \exp\left[i
\sum_{a=1}^D u_a \left(\lambda_a - \uvec{e}_a\cdot\cvh\cdot\uvec{e}_a\right) %\right.\nonumber\\
% \left.
- i\sum^D_{\substack{b,c=1 \\ b < c}} v_{\alpha} \uvec{e}_b\cdot\cvh\cdot\uvec{e}_c
\right], \nonumber \\
\eea
where the second sum is over all $L$ distinct pairs of eigenvectors ($b<c$).  Inserting the definition for $\cvh$ (eq.~\eqref{eq:ctensor}) and interchanging $\sum_{\mu}$ and $\sum_a$ in the exponential, $\Gamma(\cv)$ can be written as
\begin{multline}
\Gamma(\cv)= \frac{2^L}{(2\pi)^K}\int d^D\!u\,e^{i\bm{u}\cdot\bm{\lambda}}\int d^{L}\!v\int d^{DN} Q \\
\exp\left[-\frac{i}{N}
\sum_{\mu=1}^N\Qv_{\mu}\cdot\left\{
\mathsf{U} + \mathsf{V}\right\}\cdot\Qv_{\mu}
\right],\label{eq:q2}
\end{multline}
where the matrices  $\mathsf{U}$ and $\mathsf{V}$ are constructed from the constraint fields $u_a$ and $v_{\alpha}$ as follows 
(represented in the basis $\{\uvec{e}_a\}$) for  $D=3$ 
\begin{align}
    \mathsf{U} &= \left(
    \begin{matrix}
    u_1 & 0 & 0 \\
    0 & u_2 & 0 \\
    0 & 0 & u_3
    \end{matrix}
    \right),&
\mathsf{V} &= \left(
    \begin{matrix}
    0 & v_1 & v_2 \\
    v_1 & 0 & v_3 \\
    v_2 & v_3 & 0
    \end{matrix}
    \right),
    \end{align}
 and we have rescaled $\bm{v}$ by a factor of 2.
 %\end{widetext}
Next, we scale the $\bm{u}$ and $\bm{v}$ integrals according to 
\begin{align}
    u_a&=\frac{\bar{u}_a}{\lambda_a}& d^D\!u&=\frac{d^D\bar{u}}{\left(\det\cv \right)}&&a=1\ldots D\\
    v_{\alpha} &= \frac{\bar{v}_{\alpha}}{\sqrt{\lambda_a\lambda_b}},& d^L\!v&=\frac{d^L\bar{v}}{\left(\det\cv \right)^{(D-1)/2}}&&\stackrel{\alpha=1\ldots L}{\scriptstyle a<b\in (1, D)}\label{eq:qbar}
\end{align}
where we recognize $\det\cv = \prod_{a=1}^D\lambda_a$. The scale factors for the $v_{\alpha}$ correspond to the $u_a$ in its  row and column. To see the scaling in the measure for $\bar{v}$, note that each $v$ picks up a factor in the denominator proportional to $\lambda$. Hence the denominator scales as $\lambda^L\sim\lambda^{D(D-1)/2}\sim (\det\cv)^{(D-1)/2}$, leading to the denominator above in eq.~\eqref{eq:qbar}.  Hence, 
\begin{align}
\Gamma(\cv)&= \frac{2^L}{(2\pi)^K\left(\det\cv\right)^{(D+1)/2}}\int d^D\!\bar{u}\,e^{i{\bar{\bm{u}}\cdot\bm{1}}}\int d^{L}\!\bar{v}\int d^{DN}\,Q\nonumber\\
&
\qquad\times\exp\left[-\frac{i}{N}\sum_{\mu=1}^N\Qv_{\mu}\cdot\left\{ {\mathsf{U}} + {\mathsf{V}}\right\}\cdot\Qv_{\mu},\right],
\label{eq:q3}
\end{align}
where $\bm{1}=(1,1,\ldots)$ and now, for $(D=3)$, $\mathsf{U} + \mathsf{V}$ is expressed in the form
\begin{align}
{\mathsf{U}} + {\mathsf{V}}&=
\left(
\begin{matrix}
\displaystyle\frac{\bar{u}_1}{\sqrt{\lambda_1^2}}  &\displaystyle \frac{\bar{v}_1}{\sqrt{\lambda_1\lambda_2}}  &\displaystyle \frac{\bar{v}_2}{\sqrt{\lambda_1\lambda_3}}  \\
 \displaystyle\frac{\bar{v}_1}{\sqrt{\lambda_1\lambda_2}}  &\displaystyle \frac{\bar{u}_2}{\sqrt{\lambda_2^2}}  &\displaystyle \frac{\bar{v}_3}{\sqrt{\lambda_2\lambda_3}}  \\
 \displaystyle \frac{\bar{v}_2}{\sqrt{\lambda_1\lambda_3}}  &\displaystyle \frac{\bar{v}_3}{\sqrt{\lambda_2\lambda_3}}  &\displaystyle \frac{\bar{u}_3}{\sqrt{\lambda_3^2}}  
 \end{matrix}
 \right). \label{eq:temp123}
\end{align}
We finally rescale $\Qv_{\mu}=\bar{\Qv}_{\mu}\cdot\bm{\Lambda}$, where $\bm{\Lambda}$ is the diagonal matrix $\Lambda_{aa}=\sqrt{\lambda}_a, \Lambda_{ab}=0 (a\neq b)$. Hence, we find 
\begin{align}
\Gamma(\cv)&= \frac{2^L}{(2\pi)^K}\,
\left(\det\cv\right)^{(N-D-1)/2} \int d^D\!\bar{u}\,e^{i\bar{\bm{u}}\cdot\bm{1}}\int d^{L}\!\bar{v}\int d^{DN}\!\bar{Q}\\
&\qquad\times
\exp\left[-\frac{i}{N}\sum_{\mu=1}^N\bar{\Qv}_{\mu}\cdot\left\{ \bar{\mathsf{U}} + \bar{\mathsf{V}}\right\}\cdot\bar{\Qv}_{\mu}\right], \nonumber\\[10truept]
&\equiv\left(\det\cv\right)^{(N-D-1)/2} \,\,{\cal I},
\label{eq:zfinal}
\end{align}
where the integral ${\cal I}$ is independent of the conformation tensor $\cv$,  and $\bar{\mathsf{U}} + \bar{\mathsf{V}}$ is equal to \eqref{eq:temp123} without the factors $1/\sqrt{\lambda_a \lambda_b}$.

%

%%%%%%%%%%%%%%%%%%%%%%%%%

\bibliographystyle{unsrt}
\bibliography{refs}

\end{document}